\pdfoutput=1
\PassOptionsToPackage{unicode}{hyperref}
\PassOptionsToPackage{hyphens}{url}
\documentclass[
]{article}
\usepackage{amsmath,amssymb}
\usepackage{lmodern}
\usepackage{setspace}
\usepackage{ifxetex,ifluatex}
\ifnum 0\ifxetex 1\fi\ifluatex 1\fi=0 % if pdftex
  \usepackage[T1]{fontenc}
  \usepackage[utf8]{inputenc}
  \usepackage{textcomp} % provide euro and other symbols
\else % if luatex or xetex
  \usepackage{unicode-math}
  \defaultfontfeatures{Scale=MatchLowercase}
  \defaultfontfeatures[\rmfamily]{Ligatures=TeX,Scale=1}
\fi
\IfFileExists{upquote.sty}{\usepackage{upquote}}{}
\IfFileExists{microtype.sty}{% use microtype if available
  \usepackage[]{microtype}
  \UseMicrotypeSet[protrusion]{basicmath} % disable protrusion for tt fonts
}{}
\makeatletter
\@ifundefined{KOMAClassName}{% if non-KOMA class
  \IfFileExists{parskip.sty}{%
    \usepackage{parskip}
  }{% else
    \setlength{\parindent}{0pt}
    \setlength{\parskip}{6pt plus 2pt minus 1pt}}
}{% if KOMA class
  \KOMAoptions{parskip=half}}
\makeatother
\usepackage{xcolor}
\IfFileExists{xurl.sty}{\usepackage{xurl}}{} % add URL line breaks if available
\IfFileExists{bookmark.sty}{\usepackage{bookmark}}{\usepackage{hyperref}}
\hypersetup{
  pdftitle={Bayesian Cumulative Probability Models for Continuous and Mixed Outcomes},
  pdfauthor={Nathan T. James, Frank E. Harrell Jr., Bryan E. Shepherd},
  hidelinks,
  pdfcreator={LaTeX via pandoc}}
\usepackage{longtable,booktabs,array}
\usepackage{calc} % for calculating minipage widths
\usepackage{etoolbox}
\makeatletter
\patchcmd\longtable{\par}{\if@noskipsec\mbox{}\fi\par}{}{}
\makeatother
\IfFileExists{footnotehyper.sty}{\usepackage{footnotehyper}}{\usepackage{footnote}}
\makesavenoteenv{longtable}
\usepackage{graphicx}
\makeatletter
\def\maxwidth{\ifdim\Gin@nat@width>\linewidth\linewidth\else\Gin@nat@width\fi}
\def\maxheight{\ifdim\Gin@nat@height>\textheight\textheight\else\Gin@nat@height\fi}
\makeatother
\setkeys{Gin}{width=\maxwidth,height=\maxheight,keepaspectratio}
\makeatletter
\def\fps@figure{htbp}
\makeatother
\usepackage{titlesec}
\usepackage{booktabs}
\usepackage{longtable}
\usepackage{array}
\usepackage{multirow}
\usepackage{wrapfig}
\usepackage{float}
\usepackage{colortbl}
\usepackage{pdflscape}
\usepackage{tabu}
\usepackage{threeparttable}
\usepackage{threeparttablex}
\usepackage[normalem]{ulem}
\usepackage{makecell}

\newcommand{\beginsupplement}{
  \setcounter{table}{0}
  \renewcommand{\thetable}{S\arabic{table}}
  \setcounter{figure}{0}
  \renewcommand{\thefigure}{S\arabic{figure}}
}
\ifluatex
  \usepackage{selnolig}  % disable illegal ligatures
\fi
\newlength{\cslhangindent}
\newlength{\csllabelwidth}
\newenvironment{CSLReferences}[2] % #1 hanging-ident, #2 entry spacing
 {% don't indent paragraphs
  \setlength{\parindent}{0pt}
  \ifodd #1 \everypar{\setlength{\hangindent}{\cslhangindent}}\ignorespaces\fi
  \ifnum #2 > 0
  \setlength{\parskip}{#2\baselineskip}
  \fi
 }%
 {}
\usepackage{calc}

\newcommand{\CSLLeftMargin}[1]{\parbox[t]{\csllabelwidth}{#1}}
\newcommand{\CSLRightInline}[1]{\parbox[t]{\linewidth - \csllabelwidth}{#1}\break}

\title{Bayesian Cumulative Probability Models for Continuous and Mixed Outcomes}
\author{Nathan T. James, Frank E. Harrell Jr., Bryan E. Shepherd}
\date{2022-01-07}

\begin{document}
\maketitle

\setstretch{1.5}
\hypertarget{abstract}{%
\subsection{Abstract}\label{abstract}}

Ordinal cumulative probability models (CPMs) -- also known as cumulative link models -- such as the proportional odds regression model are typically used for discrete ordered outcomes, but can accommodate both continuous and mixed discrete/continuous outcomes since these are also ordered. Recent papers describe ordinal CPMs in this setting using non-parametric maximum likelihood estimation. We formulate a Bayesian CPM for continuous or mixed outcome data. Bayesian CPMs inherit many of the benefits of frequentist CPMs and have advantages with regard to interpretation, flexibility, and exact inference (within simulation error) for parameters and functions of parameters. We explore characteristics of the Bayesian CPM through simulations and a case study using HIV biomarker data. In addition, we provide the package \texttt{bayesCPM} which implements Bayesian CPM models using the \texttt{R} interface to the Stan probabilistic programing language. The Bayesian CPM for continuous outcomes can be implemented with only minor modifications to the prior specification and, despite some limitations, has generally good statistical performance with moderate or large sample sizes.

\hypertarget{introduction}{%
\section{1. Introduction}\label{introduction}}

Cumulative probability models for ordinal outcomes -- traditionally denoted cumulative link models {[}\protect\hyperlink{ref-agresti_categorical_2002}{1}{]} -- have been discussed extensively in the literature using both classical (frequentist) and Bayesian implementations. Since these models are characterized by adding probabilities, not link functions, we prefer the nomenclature cumulative probability model (CPM). Under the frequentist paradigm, Walker and Duncan {[}\protect\hyperlink{ref-walker_estimation_1967}{2}{]} and McCullagh {[}\protect\hyperlink{ref-peter_mccullagh_regression_1980}{3}{]} described these models as an extension of dichotomous outcome regression models such as logistic and probit regression. A Bayesian CPM for ordinal regression was explored by Albert and Chib {[}\protect\hyperlink{ref-albert_bayesian_1993}{4},\protect\hyperlink{ref-albert_bayesian_1997}{5}{]} and Johnson and Albert {[}\protect\hyperlink{ref-johnson_ordinal_1999}{6}{]}. Additional Bayesian CPM extensions including partial proportional odds {[}\protect\hyperlink{ref-peterson_partial_1990}{7}{]}, mixture link models {[}\protect\hyperlink{ref-lang_bayesian_1999}{8}{]}, location-scale ordinal regression and multivariate ordinal outcomes are described by Congdon {[}\protect\hyperlink{ref-congdon_bayesian_2005}{9}{]}. In all these settings, the number of ordered outcome categories is implicitly assumed to be much smaller than the sample size. However, continuous data where each distinct value is its own category are also ordinal and can therefore be fit using CPMs.

In the continuous outcome setting, Liu et al. {[}\protect\hyperlink{ref-liu_modeling_2017}{10}{]} demonstrate the equivalence between CPMs and semiparametric linear transformation models of the form:
\begin{equation}
Y=H(\boldsymbol{\beta}^{T}X+\varepsilon) \quad \text{with} \quad \varepsilon \sim F_{\varepsilon}
\end{equation}
where \(H(\cdot)\) is an increasing function, \(\boldsymbol{\beta}\) a vector of regression coefficients, \(X\) a vector of covariates, and \(\varepsilon\) are errors distributed according to known \(F_{\varepsilon}\). Harrell {[}\protect\hyperlink{ref-harrell_regression_2015}{11}{]}, Liu et al. {[}\protect\hyperlink{ref-liu_modeling_2017}{10}{]}, and Tian et al. {[}\protect\hyperlink{ref-tian_empirical_2019}{12}{]} describe non-parametric maximum likelihood estimation (NPMLE) {[}\protect\hyperlink{ref-zeng_maximum_2007}{13}{]} for the unspecified transformation \(H(\cdot)\) and \(\boldsymbol{\beta}\) parameters. The models have several favorable characteristics including invariance to monotonic outcome transformations for the regression coefficient estimates and the ability to handle mixed continuous and discrete outcomes such as those that arise from a lower or upper limit of detection. In addition CPMs directly model the full conditional cumulative distribution function (CDF); this allows estimates of conditional means, quantiles, and other statistics to be calculated from a single model fit. Further, because only the \(H(\cdot)\) part of the model is nonparametric, CPMs are semiparametric regression models which balance the robustness of fully nonparametric models and the efficiency of fully parametric models.

There is an extensive literature on Bayesian semiparametric regression models. The aim stated by Gelfand {[}\protect\hyperlink{ref-gelfand_approaches_1999}{14}{]} in his discussion of general approaches for these models is, ``to enrich the class of standard parametric hierarchical models by wandering nonparametrically near (in some sense) the standard class but retaining the linear structure.'\,' For example, Brunner {[}\protect\hyperlink{ref-brunner_bayesian_1995}{15}{]} describes Bayesian linear regression models with symmetric unimodal error densities and Kottas and Gelfand {[}\protect\hyperlink{ref-kottas_bayesian_2001}{16}{]} describe Bayesian semiparametric median regression.
DeYoreo \& Kottas {[}\protect\hyperlink{ref-deyoreo_bayesian_2020}{17}{]} explore Bayesian nonparametric density regression for ordinal responses by modeling the joint density between the outcome and covariates using latent continuous random variables. In the context of transformation models, Song \& Lu {[}\protect\hyperlink{ref-song_semiparametric_2012}{18}{]} develop a semiparametric transformation nonlinear mixed model which estimates the transformation, \(H(\cdot)\), and also incorporates possible nonlinear relationships between \(X\) and \(\beta\) as well as random effects using Bayesian P-splines. Tang et al. {[}\protect\hyperlink{ref-tang_semiparametric_2018}{19}{]} describe semiparametric Bayesian analysis for transformation linear mixed models using a similar Bayesian P-spline approach to estimate the transformation with a focus on nonparametric estimation of random effects. For survival outcomes, Mallick \& Walker {[}\protect\hyperlink{ref-mallick_bayesian_2003}{20}{]} describe a linear transformation model for mean survival time where the transformation, \(H(\cdot)\) and the error distribution, \(F_{\varepsilon}\), are estimated nonparametrically using mixtures of incomplete beta functions and a Pólya tree distribution, respectively. Lin et al. {[}\protect\hyperlink{ref-lin_semiparametric_2012}{21}{]} detail a semiparametric Bayesian transformation model for median survival. Hanson and colleagues {[}\protect\hyperlink{ref-damien_surviving_2013}{22},\protect\hyperlink{ref-hanson_bayesian_2007}{23}{]} and Ibrahim et al. {[}\protect\hyperlink{ref-ibrahim_bayesian_2010}{24}{]} describe other Bayesian nonparametric survival models. Additional details on general Bayesian nonparametric models can be found in the texts by Müller et al. {[}\protect\hyperlink{ref-muller_bayesian_2015}{25}{]} and Hjort et al. {[}\protect\hyperlink{ref-hjort_bayesian_2010}{26}{]}.

In this paper, we develop Bayesian CPMs for continuous and mixed outcomes.
They are distinguished from other Bayesian semiparametric approaches by their use of a simpler parametric prior specification. Bayesian CPMs inherit many of the properties of CPMs estimated using NPMLE and have additional benefits: interpretation using posterior probabilities, inference for quantities of interest without using asymptotic approximations, and the ability to incorporate available prior information.
A primary challenge when implementing Bayesian CPMs for continuous outcomes is the specification of priors for the intercept parameters used to estimate \(H(\cdot)\) and we describe several proposed strategies. Through simulations, we explore characteristics of Bayesian CPMs using several model specifications and prior combinations. A case study of HIV biomarker data with outcomes that are both right-skewed and censored at a lower limit of detection provides a real-world example. We conclude with a discussion, including advantages, current limitations, and potential extensions, and provide some recommendations for using Bayesian CPMs.

\hypertarget{methods}{%
\section{2. Methods}\label{methods}}

\hypertarget{cumulative-probability-model-formulation}{%
\subsection{Cumulative Probability Model Formulation}\label{cumulative-probability-model-formulation}}

Let \(Y_i\) be the outcome for unit \(i=1,\ldots,n\) with \(p\) covariates \(\boldsymbol{X_i}=(X_{i1},\ldots,X_{ip})\) such that each \(Y_i\) falls into one of \(j=1,\ldots, J\) ordered categories. The \(Y_i\) can be modeled using a \(Categorical(\boldsymbol{\pi_i})\) -- or \(Multinomial(1,\boldsymbol{\pi_i})\) -- distribution where \(\boldsymbol{\pi_i}=(\pi_{i1}, \ldots, \pi_{iJ})\) are the probabilities of unit \(i\) being in category \(j\) and \(\sum_{j=1}^{J}\pi_{ij}=1\). The value of \(\pi_{ij}\) is dependent on \(\boldsymbol{x_i}\), but we suppress the conditional notation for clarity. The cumulative probability of falling into category \(j\) or a lower category is \(Pr(Y_i \le j)=\eta_{ij}=\sum_{k=1}^{j}\pi_{ik}\). The CPM relates the cumulative probabilities to the observed covariates through a monotonically increasing link function \(G^{-1}(\eta_{ij})=\gamma_{j}-\boldsymbol{x_i'\beta}\). Common choices for the link function are logit, \(G^{-1}(p)=\log\left(\frac{p}{1-p}\right)\); probit, \(G^{-1}(p)=\Phi^{-1}(p)\) where \(\Phi^{-1}(p)\) is the quantile function for a standard normal distribution; and loglog, \(G^{-1}(p)=-\log(-\log(p))\). For observed data \(\{y_i,\boldsymbol{x_i}\}\) the model can be expressed as
\begin{gather}
Pr(y_i \le j|\boldsymbol{x_i},\boldsymbol{\beta},\boldsymbol{\gamma})=\eta_{ij}=G(\gamma_{j}-\boldsymbol{x_i'\beta}),
\end{gather}
where the \(\gamma_j\) are ordered continuous intercept parameters \(-\infty \equiv \gamma_0 < \gamma_1 < \cdots < \gamma_{J-1} <\gamma_J \equiv \infty\), \(\boldsymbol{\beta}\) is a vector of \(p\) coefficients, and the function \(G(\cdot)\) is a CDF defined as the inverse of the link function: standard logistic, standard normal, and standard Gumbel for the logit, probit, and loglog links, respectively. For identifiability, the linear predictor \(\boldsymbol{x_i'\beta}\) does not include an intercept. The conditional probabilities of category membership are
\begin{gather}
\label{eq:cellprobs}
\pi_{ij}=\eta_{i,j}-\eta_{i,j-1}=G(\gamma_j-\boldsymbol{x_i'\beta})-G(\gamma_{j-1}-\boldsymbol{x_i'\beta})
\end{gather}
The likelihood for an independent and identically distributed sample of outcomes \(\boldsymbol{y}=(y_1,\ldots,y_n)\) with corresponding covariates \(\boldsymbol{x}=(\boldsymbol{x_1},\ldots,\boldsymbol{x_n})\) is
\begin{gather}
p(\boldsymbol{y}|\boldsymbol{x},\boldsymbol{\gamma},\boldsymbol{\beta})=
\prod_{j=1}^{J}\prod_{i:y_i=j}[G(\gamma_j-\boldsymbol{x_i'\beta})-G(\gamma_{j-1}-\boldsymbol{x_i'\beta})]
\end{gather}
For continuous data with no ties \(J=n\); letting \(r(y_i)\) be the rank of \(y_i\), the likelihood reduces to
\begin{gather}
p(\boldsymbol{y}|\boldsymbol{x},\boldsymbol{\gamma},\boldsymbol{\beta})=
\prod_{i=1}^{n}[G(\gamma_{r(y_i)}-\boldsymbol{x_i'\beta})-G(\gamma_{r(y_i)-1}-\boldsymbol{x_i'\beta})]
\end{gather}
To complete the model specification we define priors for the parameters \(p(\boldsymbol{\beta},\boldsymbol{\gamma})\). We assume a priori independence between \(\boldsymbol{\beta}\) and \(\boldsymbol{\gamma}\) so \(p(\boldsymbol{\beta},\boldsymbol{\gamma})=p(\boldsymbol{\beta})p(\boldsymbol{\gamma})\). To simplify the model formulation we also assume noninformative priors for the regression coefficients, \(p(\boldsymbol{\beta}) \propto \boldsymbol{1}\); however weakly informative or informative priors can also be used.

Specifying priors for \(\boldsymbol{\gamma}\) is more challenging because of the ordering restriction and dimensionality. Several approaches have been suggested in the traditional CPM setting where \(J \ll n\). McKinley et al. {[}\protect\hyperlink{ref-mckinley_bayesian_2015}{27}{]} and Congdon {[}\protect\hyperlink{ref-congdon_bayesian_2005}{9}{]} describe a sequentially truncated prior distribution: \(p(\boldsymbol{\gamma})=p(\gamma_1)\prod_{j=2}^{J-1}p(\gamma_i|\gamma_{j-1})\) where \(\gamma_1 \in \mathbb{R}\) and the support of \(\gamma_j\) for \(j=2,\ldots, J-1\) is \((\gamma_{j-1},\infty)\). For example using normal and truncated normal priors, \(p(\gamma_1)\sim N(0, \sigma_\gamma^2)\) and \(p(\gamma_j|\gamma_{j-1}) \sim N(0, \sigma_\gamma^2)I(\gamma_{j-1},\infty)\). A second approach described by Albert and Chib {[}\protect\hyperlink{ref-albert_bayesian_1997}{5}{]} defines the prior on a transformation of the intercepts to an unconstrained space; first normalizing \(\gamma_0\) to 0 so \(0 \equiv \gamma_0 < \gamma_1 < \cdots < \gamma_{J-1} <\gamma_J \equiv \infty\) and then letting \(\delta_1=\log(\gamma_1)\) and \(\delta_j=\log(\gamma_j - \gamma_{j-1}),\, 2 \le j \le J-1\) a multivariate prior can be assigned, e.g.~\(\boldsymbol{\delta} \sim N_{J-1}(\boldsymbol{\mu_0},\boldsymbol{\Sigma_0})\).
Both approaches provide priors that satisfy the ordering restriction, but may be cumbersome when the number of distinct categories is high. The first requires specification of the distribution and its hyperparameters, then sampling from the sequential series of \(J-2\) truncated distributions; the second requires specification of the \(J-1\) dimensional \(\boldsymbol{\mu_0}\) vector and the \(J-1 \times J-1\) dimensional covariance matrix \(\boldsymbol{\Sigma_0}\).

We instead adopt a third approach which defines a prior on \(\boldsymbol{\pi_i}\) for a prespecified covariate vector and utilizes the transformation defined by \(G(\cdot)\) to induce a prior on \(\boldsymbol{\gamma}\) {[}\protect\hyperlink{ref-betancourt_ordinal_2019}{28}{]}.
Let \(\pi_{.j} \equiv Pr(r(y)=j|\boldsymbol{x}=0)\) be the probability of being in category \(j\) when all covariates are 0 and \(\boldsymbol{\pi_{.}}=(\pi_{.1},\ldots,\pi_{.J})\).
It may be useful to center the covariates by using \(x'=x-\bar{x}\) in place of \(x\). Then \(\pi_{.j}\) is the probability of being in category \(j\) when all covariates are at their mean value. From equation (\ref{eq:cellprobs}) it follows that
\begin{gather}
\pi_{.j}=G(\gamma_j-0)-G(\gamma_{j-1}-0)=G(\gamma_j)-G(\gamma_{j-1})
\end{gather}
These equations define a transformation \(h(\boldsymbol{\gamma})=\boldsymbol{\pi_{.}}\) between the intercept parameters and probabilities of category membership when \(\boldsymbol{X}=\boldsymbol{0}\). Conversely,
\begin{gather}
\label{eq:invtrans}
\sum_{k=1}^{j}\pi_{.k}=\sum_{k=1}^{j}\left[G(\gamma_k)-G(\gamma_{k-1})\right]=G(\gamma_j)
\end{gather}
so \(G^{-1}\left(\sum_{k=1}^{j}\pi_{.k}\right)=\gamma_j\) defines the inverse transformation \(h^{-1}(\boldsymbol{\pi_{.}})=\boldsymbol{\gamma}\). Because \(\boldsymbol{y}\) has a multinomial distribution a conjugate Dirichlet distribution with hyperparameters \(\boldsymbol{\alpha}\) is a natural choice of prior for \(\boldsymbol{\pi_{.}}\). Setting \(p(\boldsymbol{\pi_{.}}|\boldsymbol{\alpha}) \propto \prod_{j=1}^{J}\pi_{.j}^{\alpha_j-1}\) the posterior distribution is
\begin{align}
p(\boldsymbol{\gamma},\boldsymbol{\beta}|\boldsymbol{x},\boldsymbol{y}) & \propto p(\boldsymbol{\gamma})p(\boldsymbol{\beta}) p(\boldsymbol{y}|\boldsymbol{x},\boldsymbol{\gamma},\boldsymbol{\beta})\\
&\propto p(h(\boldsymbol{\gamma}))|\mathcal{J}|p(\boldsymbol{\beta}) p(\boldsymbol{y}|\boldsymbol{x},\boldsymbol{\gamma},\boldsymbol{\beta})\\
\label{eq:post}
&\propto p(\boldsymbol{\pi_{\cdot}}|\boldsymbol{\alpha})|\mathcal{J}|p(\boldsymbol{\beta}) p(\boldsymbol{y}|\boldsymbol{x},\boldsymbol{\gamma},\boldsymbol{\beta})
\end{align}
where \(\mathcal{J}\) is the Jacobian of the transformation \(h(\boldsymbol{\gamma})=\boldsymbol{\pi_{.}}\). Letting \(\Omega=\sum_{j=1}^J\pi_{.j}=1\) be the constraint that all category probabilities sum to 1, the entries in \(\mathcal{J}\) where \(\mathcal{J}_{r,c}\) is the term in row \(r\) and column \(c\) are
\begin{equation*}
\mathcal{J}_{j,1}=\frac{\partial \pi_{.j}}{\partial \Omega}=1, \quad
\mathcal{J}_{j+1,j+1}=\frac{\partial \pi_{.j+1}}{\partial \gamma_j}=\frac{\partial}{\partial \gamma_j}\left[G(\gamma_{j+1})-G(\gamma_{j})\right]=-g(\gamma_j), \quad
\mathcal{J}_{j,j+1}=\frac{\partial \pi_{.j}}{\partial \gamma_j}=\frac{\partial}{\partial \gamma_j}\left[G(\gamma_{j})-G(\gamma_{j-1})\right]=g(\gamma_j),
\end{equation*}
where \(j=1,\ldots,J-1\), \(g(\cdot)\) is the density function of the distribution \(G(\cdot)\), and \(\mathcal{J}_{r,c}=0\) for all other entries; the form of the Jacobian is

\begin{equation}
\begin{vmatrix}
1      &  g(\gamma_1)  &  0            & 0           & \cdots           & 0 \\
1      & -g(\gamma_1)  &  g(\gamma_2)  & 0           & \cdots           & 0 \\
1      & 0             & -g(\gamma_2)  & g(\gamma_3) & \cdots           & 0 \\
\vdots & \vdots        &  \vdots       & \vdots      & \ddots           & \vdots \\
1      & 0             &  0            & 0           & -g(\gamma_{j-1}) & g(\gamma_j)\\
1      & 0             &  0            & 0           & 0                & -g(\gamma_j)\\
\end{vmatrix}
\end{equation}

\begin{figure}

{\centering \includegraphics[width=0.95\linewidth]{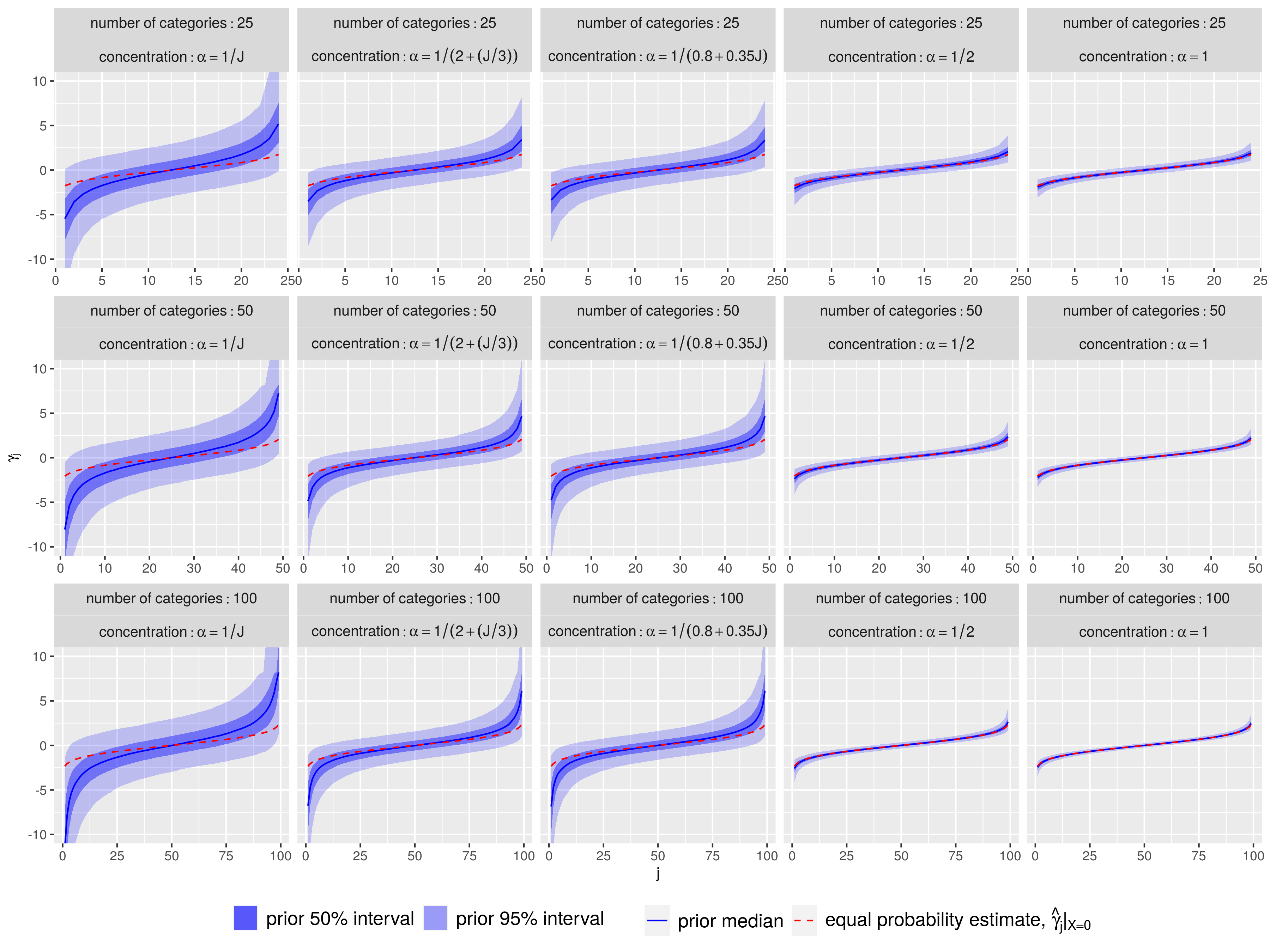} 

}

\caption{Induced $\boldsymbol{\gamma}$ priors under probit link, $G^{-1}(\cdot)=\Phi^{-1}(\cdot)$. Each subplot displays the median and credible intervals of $\gamma_j$ for $j=1,\ldots,J-1$}\label{fig:probit-induced}
\end{figure}

While it is possible to define separate \(\alpha_j\) parameters for each category, we restrict our attention to symmetric Dirichlet distributions which use a single \(\alpha\) value for all categories (i.e., \(\alpha_1 = \alpha_2 = \cdots = \alpha_J\)) so \(\boldsymbol{\alpha}=\alpha\boldsymbol{1}\) where \(\boldsymbol{1}\) is a \(J-1\) dimensional vector of 1s. The symmetric Dirichlet prior on \(\boldsymbol{\pi_{.}}\) along with the inverse transformation \(h^{-1}(\cdot)\) defined in equation (\ref{eq:invtrans}) induces a prior for \(\boldsymbol{\gamma}\) with \(\boldsymbol{\alpha}\) controlling the concentration of the induced prior. For example, Figure \ref{fig:probit-induced} shows induced \(\boldsymbol{\gamma}\) priors assuming a probit link for several combinations of concentration parameter and number of categories. The priors are approximately distributed around the intercepts that result under an assumption of equal probability for all categories when \(\boldsymbol{X}=\boldsymbol{0}\); that is, for \(J=n\) the values \(G^{-1}(\sum_{k=1}^j 1/n)=G^{-1}(j/n)=\hat{\gamma}_{j|X=0}\). The prior choices correspond to several options for a multinomial-Dirichlet model {[}\protect\hyperlink{ref-gelman_bayesian_2014}{29}{]}: a uniform Dirichlet (\(\boldsymbol{\alpha}=1\)), the multivariate Jeffreys prior (\(\boldsymbol{\alpha}=1/2\)), an overall objective prior recommended by Berger et al. {[}\protect\hyperlink{ref-berger_overall_2015}{30}{]} (\(\boldsymbol{\alpha}=1/J\)), and two additional `reciprocal' priors (\(\boldsymbol{\alpha}=1/(2+(J/3))\) and \(\boldsymbol{\alpha}=1/(0.8+0.35J)\)). The last two priors were found using a trial-and-error procedure in a simulation study with the aim of minimizing the difference between the posterior mean and mode intercept estimates and the corresponding maximum likelihood intercept estimates. As the number of categories increases the uniform and Jeffreys priors more strongly favor intercepts assuming equal probability for all categories; in contrast, the three reciprocal priors are adjusted to maintain the same degree of concentration relative to the equal probability intercepts, \(\hat{\gamma}_{j|X=0}\).

In multiparameter models, the choice of an objective reference prior depends on the parameter or statistic of interest (e.g.~\(\boldsymbol{\beta}\), conditional CDF, conditional mean) {[}\protect\hyperlink{ref-berger_overall_2015}{30}{]}. Without prior information, we seek a value of \(\boldsymbol{\alpha}\) with minimal impact on inference for a variety of settings and quantities of interest while still producing posterior estimates that can be sampled well by the MCMC algorithm.

\hypertarget{estimation}{%
\subsubsection{Estimation}\label{estimation}}

The model in (\ref{eq:post}) is implemented using the \texttt{R} interface to Stan {[}\protect\hyperlink{ref-stan_development_team_rstan:_2018}{31}{]} which performs MCMC sampling for the posterior parameters using no-U-Turn Hamiltonian Monte Carlo {[}\protect\hyperlink{ref-gelman_bayesian_2014}{29},\protect\hyperlink{ref-neal_mcmc_2011}{32}{]}. The \texttt{R} package \texttt{bayesCPM} which implements the Bayesian CPM model described in this paper is available through github at \url{https://github.com/ntjames/bayesCPM/tree/master/pkg}.

\hypertarget{posterior-conditional-quantities}{%
\subsubsection{Posterior Conditional Quantities}\label{posterior-conditional-quantities}}

Using the \(S\) draws from the posterior distribution, (\(\tilde{\boldsymbol{\gamma}}^{(s)}\), \(\tilde{\boldsymbol{\beta}}^{(s)}\)) where \(s=1,\ldots,S\), it is straightforward to calculate the distribution of the posterior conditional CDF, mean, quantiles or other functions of the parameters. For example, the distribution of the posterior conditional CDF at \(y_j\) with covariates \(\boldsymbol{x}\) can be approximated by the \(S\) values \(\tilde{F}^{(s)}(y_j|\boldsymbol{x})=G(\tilde{\gamma}_{r(y_j)}^{(s)}-\boldsymbol{x}^{T}\tilde{\boldsymbol{\beta}}^{(s)})\) and the complete conditional CDF can be obtained by a step function connecting \(\tilde{F}^{(s)}(y_j|\boldsymbol{x})\) for \(j=1,\ldots,J\). The posterior mean distribution conditional on covariate vector \(\boldsymbol{x}\) is approximated by \(\tilde{E}^{(s)}[Y|\boldsymbol{x}]=\sum_{j=1}^{J}y_j\tilde{f}^{(s)}(y_j|\boldsymbol{x})\) where \(\tilde{f}^{(s)}(y_j|\boldsymbol{x})=\tilde{F}^{(s)}(y_j|\boldsymbol{x})-\tilde{F}^{(s)}(y_{j-1}|\boldsymbol{x})\) and \(\tilde{F}^{(s)}(y_0|\boldsymbol{x}) \equiv 0\) so \(\tilde{f}^{(s)}(y_1|\boldsymbol{x})=\tilde{F}^{(s)}(y_1|\boldsymbol{x})\). Note that for mixed continuous/discrete outcomes, such as those arising from a detection limit, data below or above the limit do not have a known \(y_j\) value; in this case a value must be assigned to calculate the conditional mean. To estimate the \(q^{th}\) posterior conditional quantile we first find \(y_j^{(s)}=\inf\{y:\tilde{F}^{(s)}(y|\boldsymbol{x})\ge q\}\) and the next smallest value \(y_{j-1}^{(s)}\), then use linear interpolation to find quantile \(y_q^{(s)}\) where \(y_{j-1}^{(s)}<y_q^{(s)}<y_j^{(s)}\). For each of these functionals, point and interval estimates can be obtained by summarizing the \(S\) values obtained from the posterior parameter draws without using asymptotic approximations. For example, the mean of the posterior conditional CDF distribution is \(\frac{1}{S}\sum_{s=1}^S\tilde{F}^{(s)}(y_j|\boldsymbol{x})\) and the 2.5\% and 97.5\% percentiles of the \(y_q^{(s)}\) values are the bounds of a 95\% credible interval for the \(q^{th}\) posterior conditional quantile.

\hypertarget{simulations}{%
\section{3. Simulations}\label{simulations}}

\hypertarget{set-up}{%
\subsection{Set-up}\label{set-up}}

To evaluate the properties of the Bayesian CPM for continuous and mixed outcomes we generate data from several simulation scenarios:
\begin{align*}
1.\; Y&=\exp(X_1\beta_1 + X_2 \beta_2 + \varepsilon) \quad \varepsilon \sim N(0,1)\\
2.\; Y&=\exp(X_1\beta_1 + X_2 \beta_2 + \varepsilon) \quad \varepsilon \sim Logistic(0,1/3)\\
3.\; Y&=X_1\beta_1 + X_2 \beta_2 + \varepsilon \quad \varepsilon \sim Gumbel(0,1)
\end{align*}
where \(\beta_1=1\), \(\beta_2=-0.5\), \(X_1 \sim Bernoulli(0.5)\) and \(X_2 \sim N(0,1)\). For each scenario a second set of simulations was used to evaluate a mixed discrete/continuous outcome with a lower limit of detection; for scenario (1) and (2) values of \(Y<1\) to were set to 1, for scenario (3) values of \(Y<0\) were set to 0. The uncensored and censored outcome data based on (1) and (3) were evaluated using a Bayesian CPM with the properly specified probit and loglog links, respectively. For scenario (2) a logit link Bayesian CPM (which implies \(\varepsilon \sim Logistic(0,1)\)) was used. In each of the six outcome models, three \(\boldsymbol{\alpha}\) concentration hyperparameters (\(1/J\), \(1/(2+(J/3))\), and \(1/(0.8+0.35J)\)) were considered for \(p(\boldsymbol{\pi_{.}}|\boldsymbol{\alpha})\) for 18 model and prior combinations. Sample sizes \(n=25,50,100,200\) and \(400\) were used under each model/prior combination for a total of 90 simulation models. \(1,000\) datasets were generated under each simulation model.

We examine the average percent bias of the posterior median for parameters \(\beta_1\) and \(\beta_2\) and five \(\gamma_j\) parameters corresponding to \(y\) values spaced across the range of the data: For scenario (1) \(\boldsymbol{y}=\{y_1=e^{-1},y_2=e^{-0.33},y_3=e^{0.5},y_4=e^{1.33},y_5=e^2\}\), for scenario (2) \(\boldsymbol{y}=\{y_1=e^{-0.5},y_2=e^{0},y_3=e^{0.5},y_4=e^{1},y_5=e^{1.5}\}\) and for scenario (3) \(\boldsymbol{y}=\{y_1=-0.3,y_2=0,y_3=0.5,y_4=1.5,y_5=2.5\}\). For the censored outcomes, estimates are only available for the values of \(\boldsymbol{y}\) above the censoring threshold. We also calculate average percent bias of the conditional CDF for \(\boldsymbol{y}\) when \(X_1=1\) and \(X_2=1\), and the conditional median, mean and 20th percentile at \((X_1=1,X_2=1)\) and \((X_1=1,X_2=0)\).

\hypertarget{results}{%
\subsection{Results}\label{results}}

A Bayesian CPM was fit to each of the 1000 simulation datasets for each scenario/prior/sample size combination. For each simulation dataset, the median of the posterior distribution of the parameter or conditional CDF, mean, or quantile was used as a point estimate. These point estimates were compared to the true value from the generating model and the results averaged over all simulation datasets. Each model was run with 2 MCMC chains using 2000 warmup and 2000 sampling iterations each. Retaining only the sampling iterations from each chain for inference resulted in a total of 4000 posterior parameter vector draws per model.

In general, the Bayesian CPM had reasonable performance in estimating parameters and conditional quantities for the simulation settings explored; especially for larger sample sizes. However, performance was poor for some quantities and may be sensitive to the conditioning covariate values and censoring threshold. The three \(\boldsymbol{\alpha}\) values produced similar results for most scenarios and no prior choice was best across all parameters and quantities of interest.

\hypertarget{parameters}{%
\subsubsection{Parameters}\label{parameters}}

\begin{figure}

{\centering \includegraphics[width=0.9\linewidth]{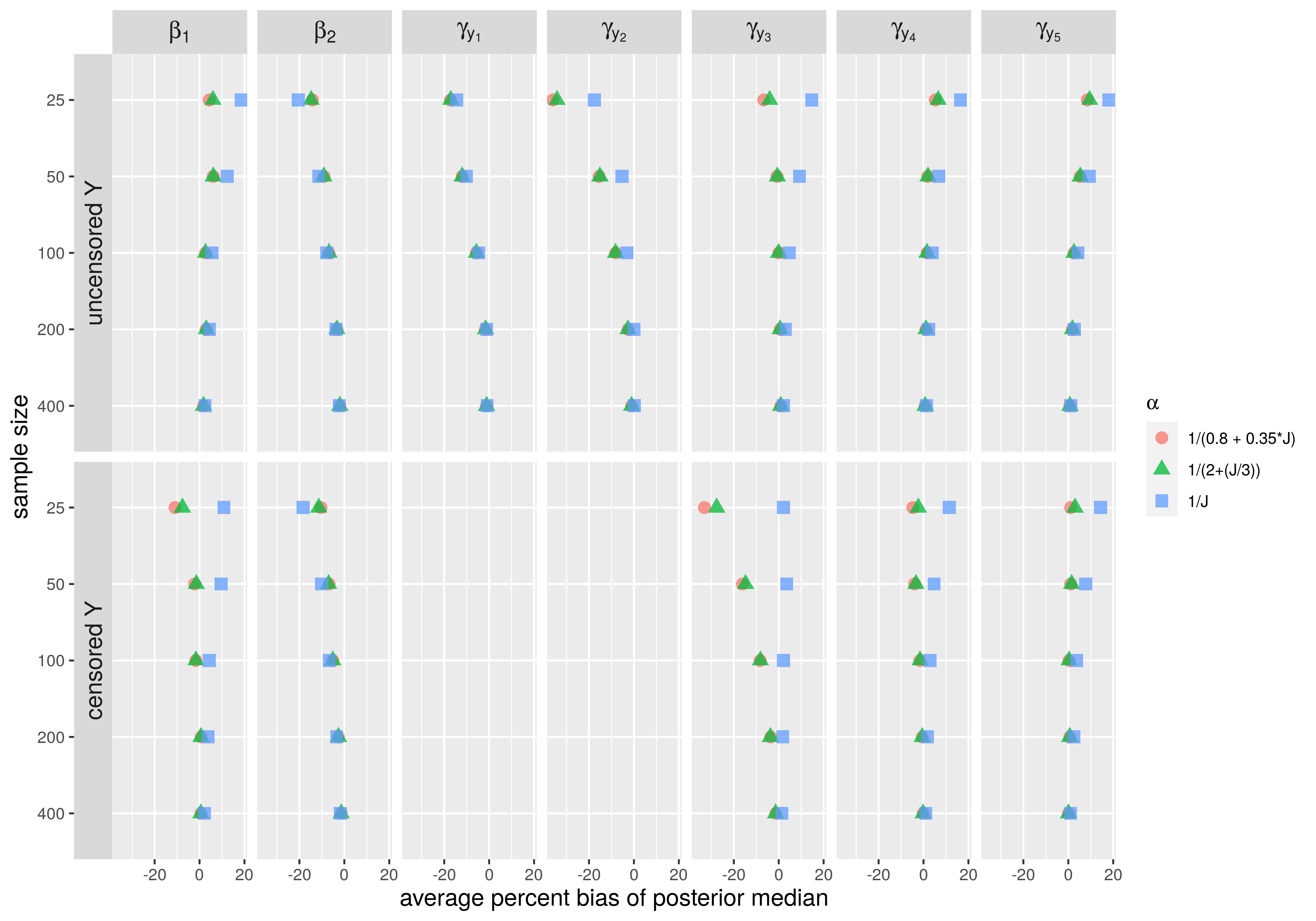} 

}

\caption{Percent bias in parameters for simulations using probit link}\label{fig:simplt-pars-1}
\end{figure}

For scenario (1) using a properly specified probit link CPM, the average percent bias in the posterior median for \(\beta_1\), \(\beta_2\), and \(\gamma_{y_k}\) is shown in Figure \ref{fig:simplt-pars-1} for the uncensored and censored outcome data. Average percent bias was largest for the smallest sample sizes, but the direction and magnitude of the bias depended on the outcome, concentration prior and parameter. Across both outcomes, the estimates of \(\beta_1\) and all \(\gamma\)s were larger using the \(\boldsymbol{\alpha}=1/J\) concentration prior than the \(\boldsymbol{\alpha}=1/(0.8+0.35J)\) or \(\boldsymbol{\alpha}=1/(2+(J/3))\) concentration priors while the \(\beta_2\) estimates were smaller with \(\boldsymbol{\alpha}=1/J\). For the \(\beta\) parameters, the priors \(\boldsymbol{\alpha}=1/(0.8+0.35J)\) and \(\boldsymbol{\alpha}=1/(2+(J/3))\) produced less biased estimates than \(\boldsymbol{\alpha}=1/J\) for both outcomes. The situation was more complex for the \(\gamma\) parameters. With the uncensored outcome, the \(\boldsymbol{\alpha}=1/J\) prior estimate was less biased for \(\gamma_{y_1}\) and \(\gamma_{y_2}\), but more biased for \(\gamma_{y_3}\), \(\gamma_{y_4}\), and \(\gamma_{y_5}\); with the censored outcome, the \(\boldsymbol{\alpha}=1/J\) prior estimate was less biased for \(\gamma_{y_3}\), but more biased for \(\gamma_{y_4}\) and \(\gamma_{y_5}\).

\begin{figure}

{\centering \includegraphics[width=0.9\linewidth]{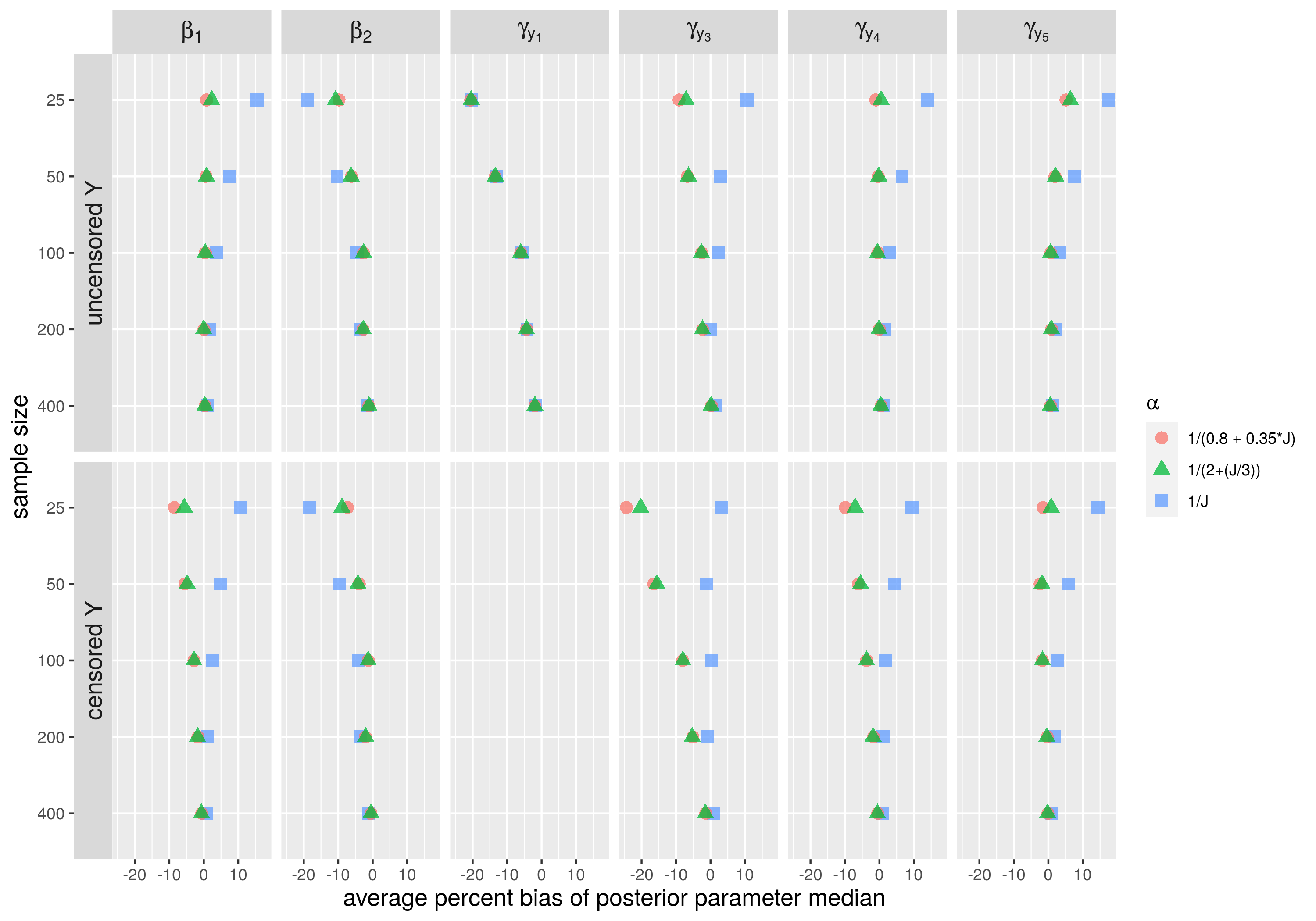} 

}

\caption{Percent bias in parameters for simulations using logit link}\label{fig:simplt-pars-2}
\end{figure}

Figure \ref{fig:simplt-pars-2} shows the average percent bias in the posterior median for \(\beta_1\), \(\beta_2\), and \(\gamma_{y_k}\) for scenario (2) using a logit link CPM. Unlike scenario (1), the assumed scale of the latent variable with logit link (\(\varepsilon \sim Logistic(0,1)\)) does not match the scale from the simulation model (\(\varepsilon \sim Logistic(0,1/3)\)). In this case it can be shown that the CPM parameter estimates are proportional to the parameters from the generating simulation model. Assume latent \(Y^*=\beta'X + a\varepsilon\) with known \(\varepsilon \sim F_{\varepsilon}\) and constant scaling factor \(a>0\), and observed \(Y=H(Y^*)\) with increasing function \(H(t)\). Then \(Y=H(\beta'X+a\varepsilon)=H'(\xi'X+\varepsilon)\) where \(\xi=a^{-1}\beta\) and \(H'(t)=H(at)\) so \(Pr(Y \le y|X)=Pr(H'(\xi'X+\varepsilon)\le y|X)=F_{\varepsilon}(H'^{-1}(y)-\xi'X)\). Using a CPM with link function \(F_{\varepsilon}^{-1}\) to analyze the observed outcome \(Y\) results in estimates of \(\xi=a^{-1}\beta\) for the linear predictor coefficients and \(H'^{-1}=a^{-1}H^{-1}\) for the intercept function. To compare the CPM model estimates (e.g.~\(\xi\), \(H'^{-1}\)) to the generating model parameters (\(\beta\), \(H^{-1}\)) it is necessary to rescale by \(a\). Conceptually this is equivalent to rescaling \(\varepsilon\) for the
latent \(Y^*\) to match the assumed scale before fitting the CPM. Outside of simulations, the scale factor is not known but can be assumed to equal 1 without loss of generality because \(Y^*\) is latent; therefore rescaling is not necessary in practice. In general, simulation results were similar to those in Figure \ref{fig:simplt-pars-1}; bias was small with moderate sample sizes.

\begin{figure}

{\centering \includegraphics[width=0.65\linewidth]{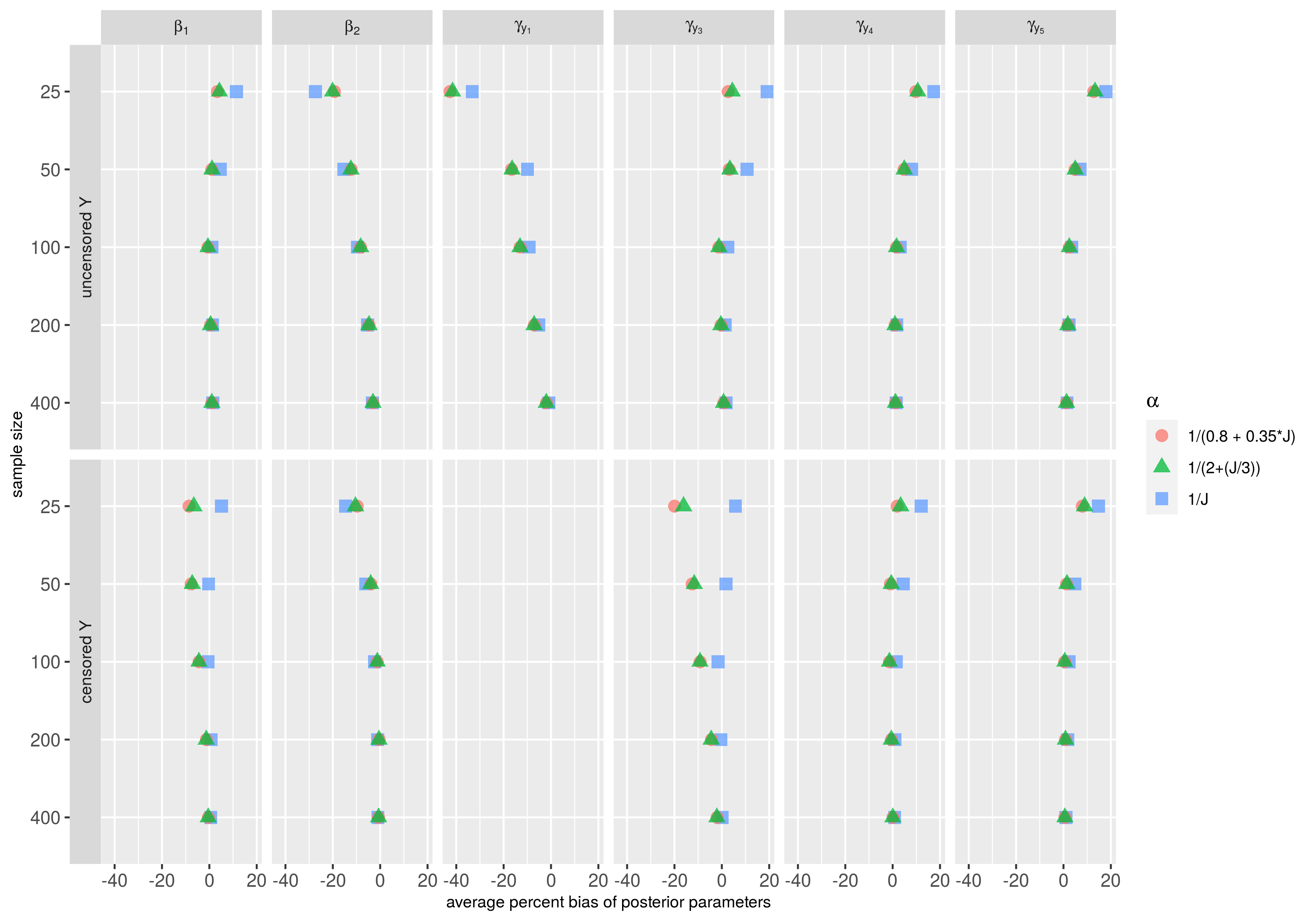} 

}

\caption{Bias in parameters for simulations using loglog link}\label{fig:simplt-pars-3}
\end{figure}

For scenario (3) using the correctly specified loglog link with an identity transformation, overall trends resembled those in scenario (1) (see Supp. Figure \ref{fig:simplt-pars-3}).

\hypertarget{conditional-cdf}{%
\subsubsection{Conditional CDF}\label{conditional-cdf}}

\begin{figure}

{\centering \includegraphics[width=0.9\linewidth]{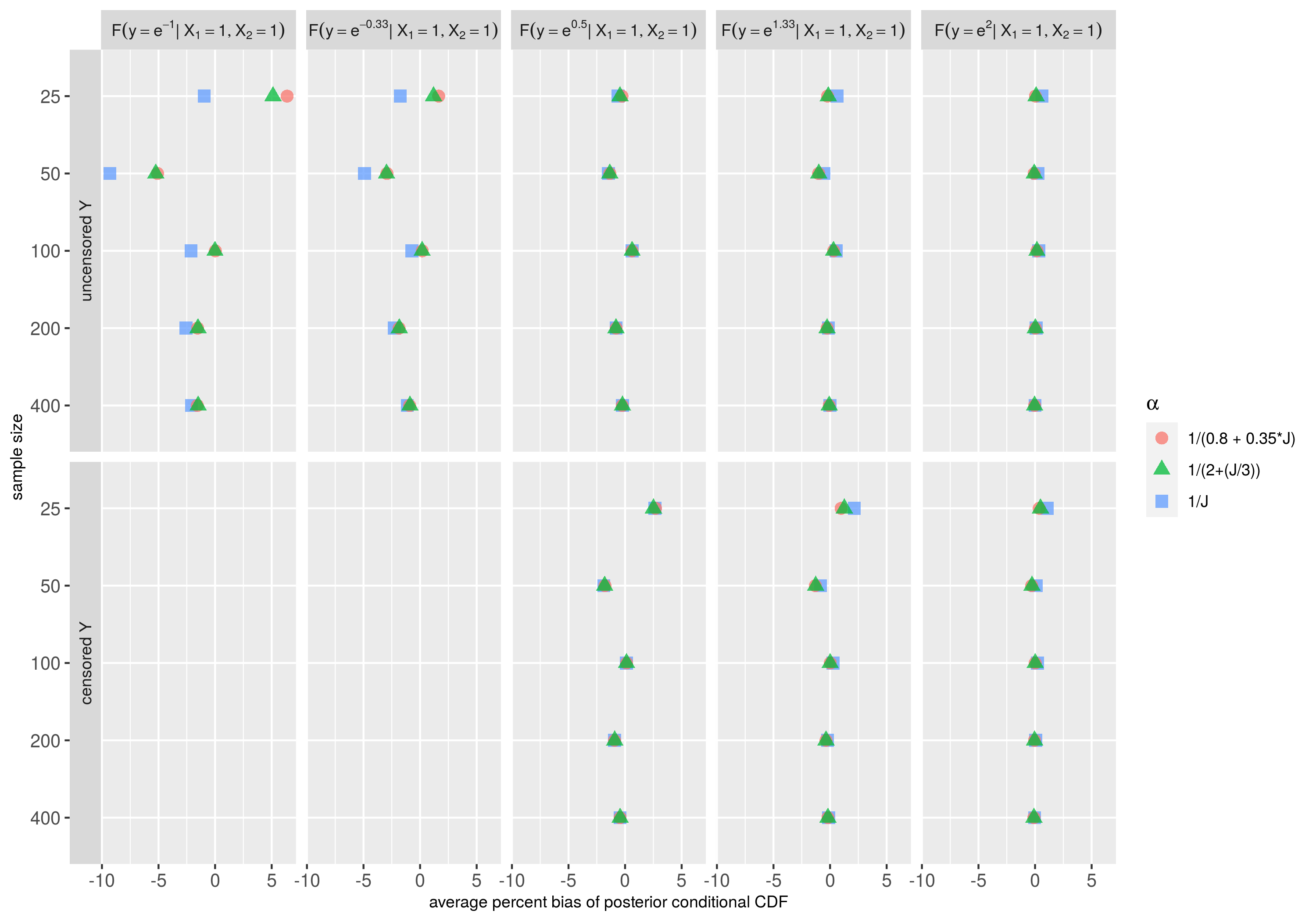} 

}

\caption{Percent bias in conditional CDF for simulations using probit link}\label{fig:simplt-cdf-1}
\end{figure}

\begin{figure}

{\centering \includegraphics[width=0.65\linewidth]{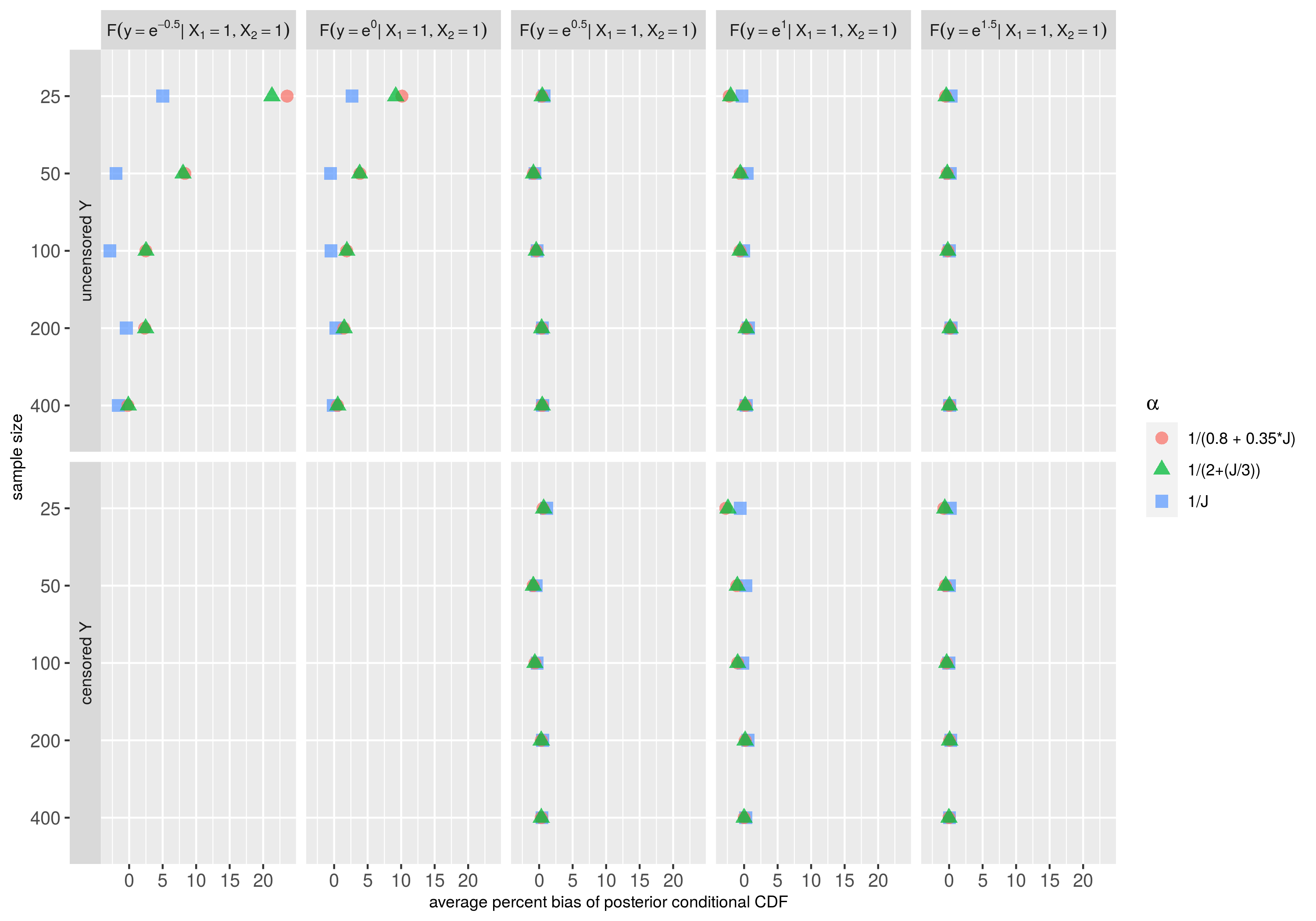} 

}

\caption{Percent bias in conditional CDF for simulations using logit link}\label{fig:simplt-cdf-2}
\end{figure}

\begin{figure}

{\centering \includegraphics[width=0.75\linewidth]{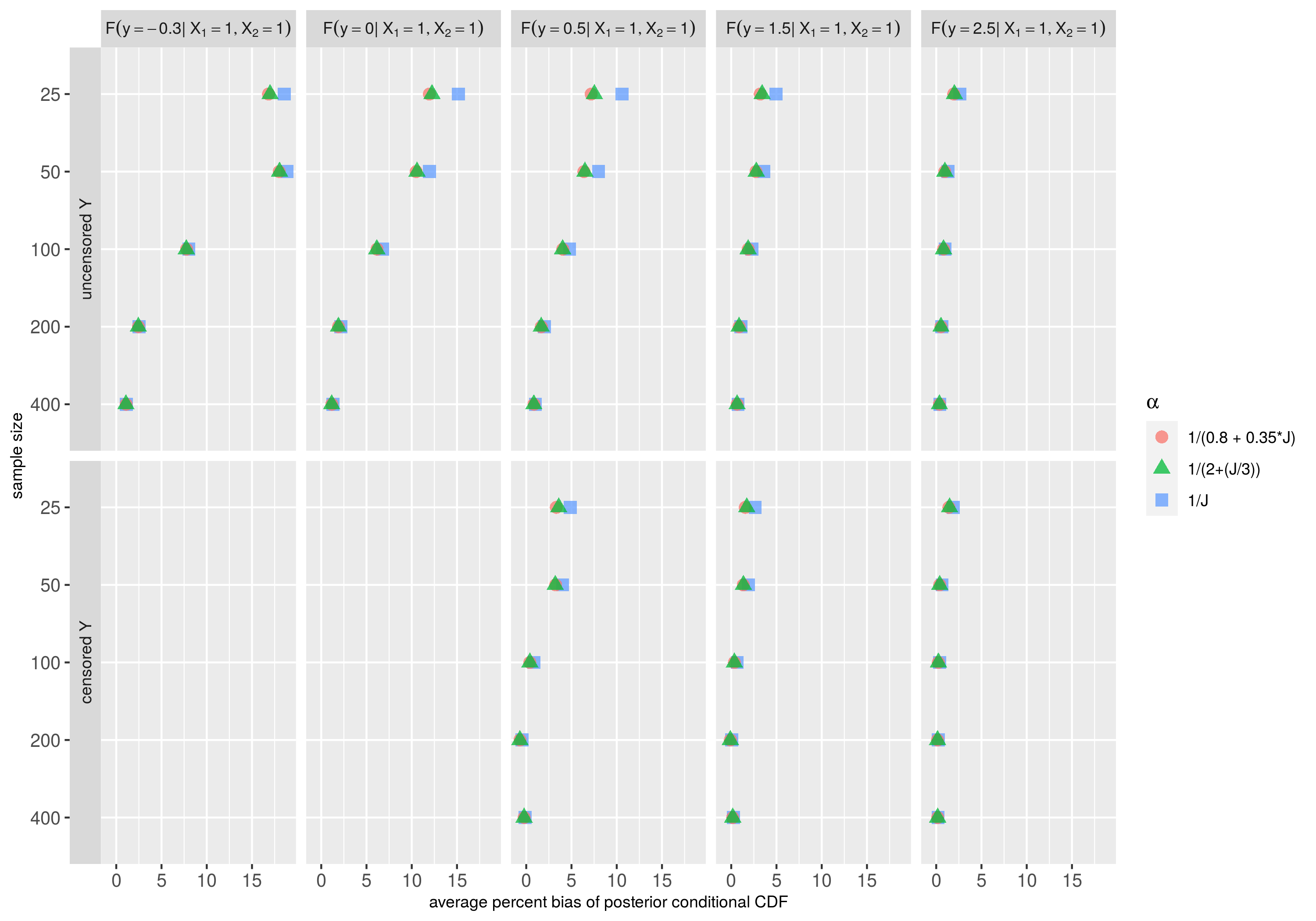} 

}

\caption{Percent bias in conditional CDF for simulations using loglog link}\label{fig:simplt-cdf-3}
\end{figure}

Figure \ref{fig:simplt-cdf-1} shows the average percent bias in the posterior conditional CDF, \(F(y|X_1=1,X_2=1)\), for scenario (1). At the values \(y_1=e^{-1},y_2=e^{-0.33},y_3=e^{0.5},y_4=e^{1.33},y_5=e^2\) the true conditional CDF values were around 0.07, 0.20, 0.50, 0.8, and 0.93, respectively. For the uncensored outcome, the conditional CDF estimates had larger percent bias when \(y<e^{0.5}\), especially for the sample sizes \(n=25\) and \(n=50\). This is not surprising, as it is difficult to estimate a conditional CDF at the tail of distribution with a small sample size. In addition, for conditional CDF estimates at \(y<e^{0.5}\), the concentration prior \(\boldsymbol{\alpha}=1/J\) produced estimates that were lower than the other reciprocal priors. The direction of the bias did not show a consistent trend across sample sizes. Similar patterns were seen for the censored outcome, less biased estimates for the CDF at higher \(y\) values and larger sample sizes. The results were much the same for scenarios (2) and (3) under both outcomes: larger average percent bias for the conditional estimates of \(F(y|X_1=1,X_2=1)\) for lower values of \(y\) and smaller sample sizes (Supp. Figures \ref{fig:simplt-cdf-2} and \ref{fig:simplt-cdf-3}).

\hypertarget{conditional-mean}{%
\subsubsection{Conditional Mean}\label{conditional-mean}}

\begin{figure}

{\centering \includegraphics[width=0.9\linewidth]{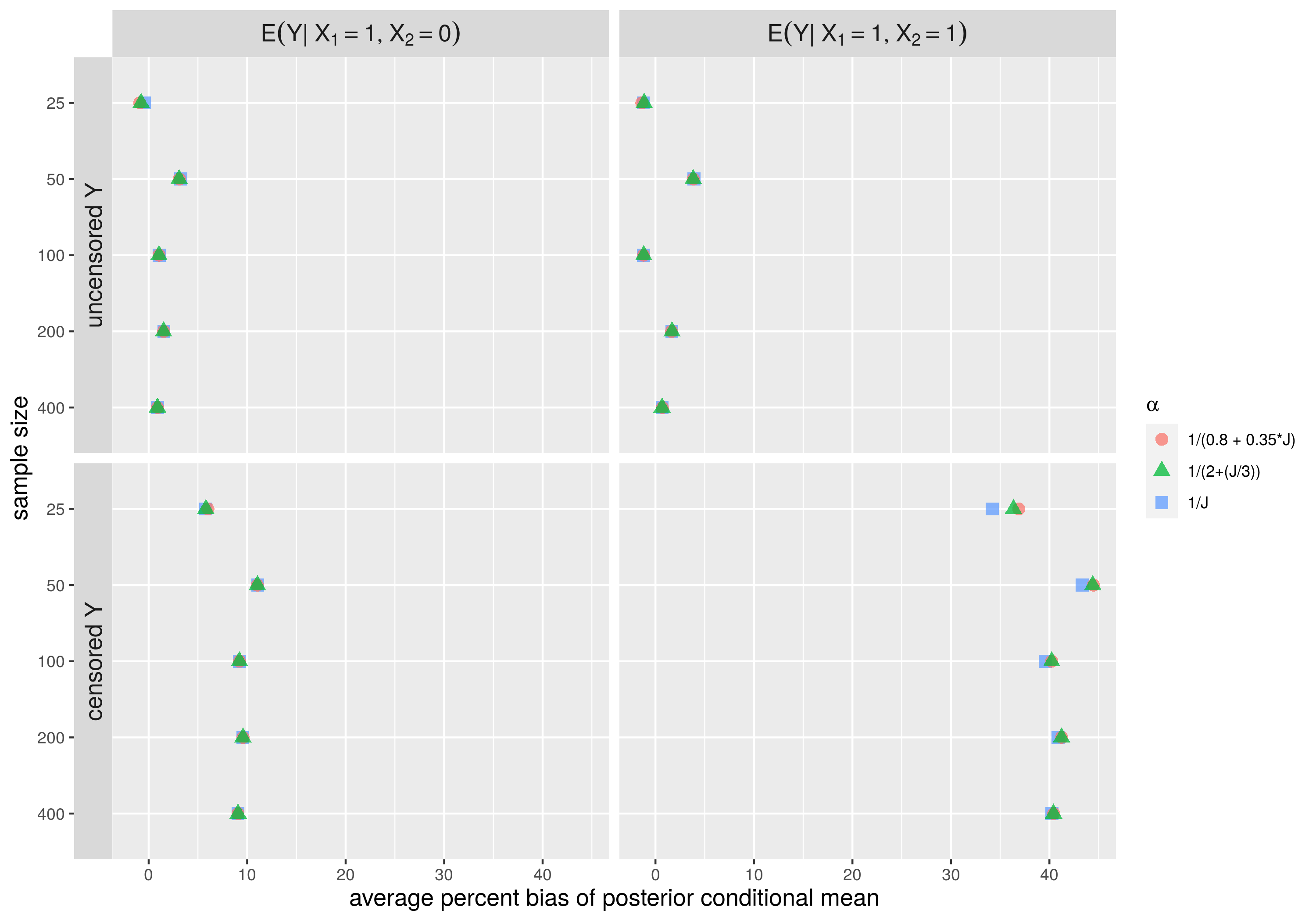} 

}

\caption{Percent bias in conditional mean for simulations using probit link}\label{fig:simplt-mn-1}
\end{figure}

\begin{figure}

{\centering \includegraphics[width=0.75\linewidth]{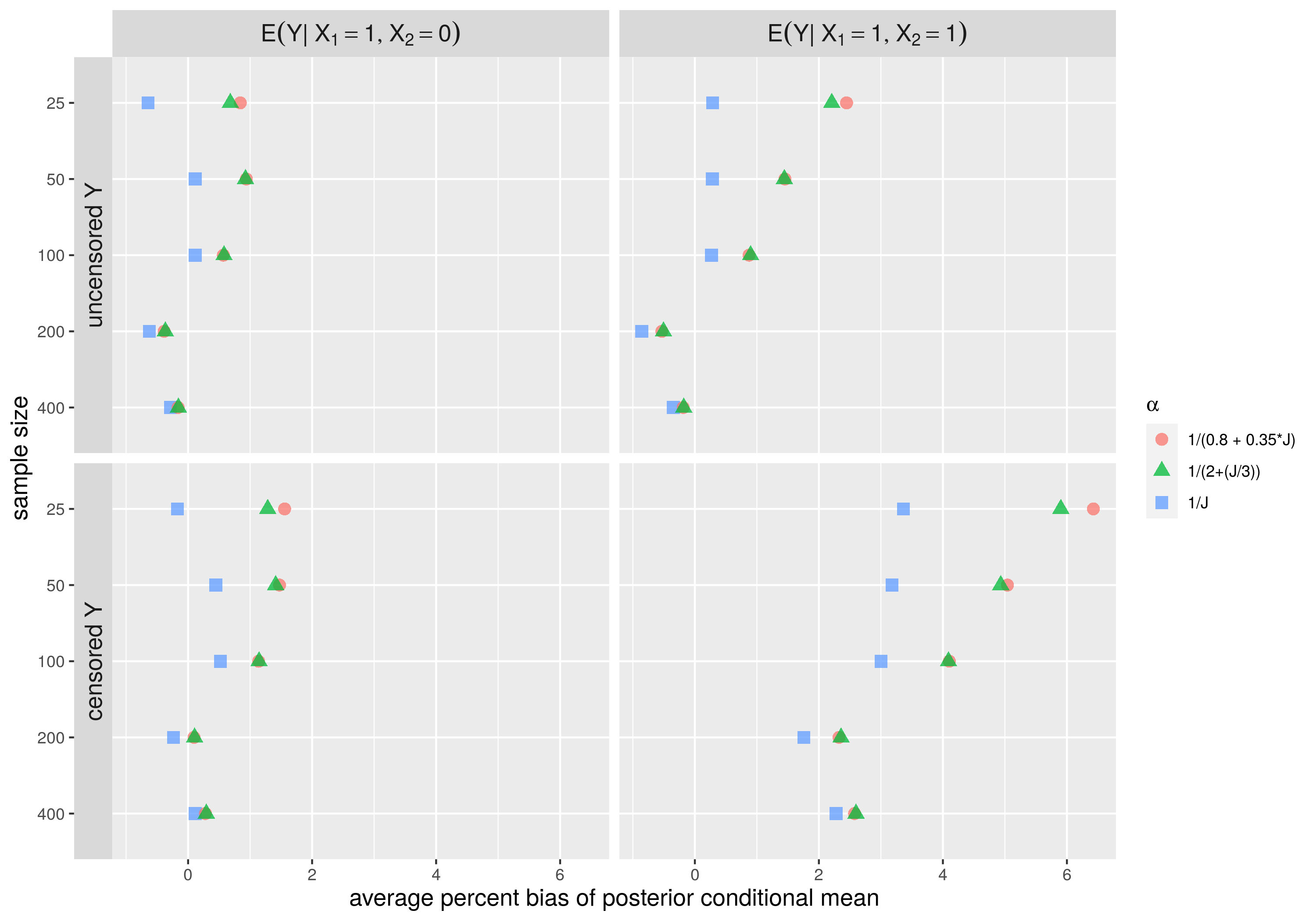} 

}

\caption{Percent bias in conditional mean for simulations using logit link}\label{fig:simplt-mn-2}
\end{figure}

The top row of Figure \ref{fig:simplt-mn-1} presents the average percent bias in the posterior conditional mean for the uncensored simulation outcomes at \((X_1=1,X_2=0)\) and \((X_1=1,X_2=1)\) in scenario (1). For this scenario, the average percent bias was less than 5\% for all sample sizes and priors. In contrast, the bottom row of Figure \ref{fig:simplt-mn-1} shows the bias in posterior conditional mean estimates for the censored outcomes where a value of \(y=1\) was used in the conditional mean calculation for outcomes censored at \(Y<1\). Using the censoring threshold value for censored observations results in inflated average percent bias compared to the uncensored case depending on where the threshold falls in relation to the true conditional distribution. For example, the average percent bias of \(E(Y|X_1=1,X_2=1)\) for the censored outcome in scenario (1) was around 40\% even for the largest sample size. Results were similar for scenario (2) (see Supp. Figure \ref{fig:simplt-mn-2}).

\begin{figure}

{\centering \includegraphics[width=0.9\linewidth]{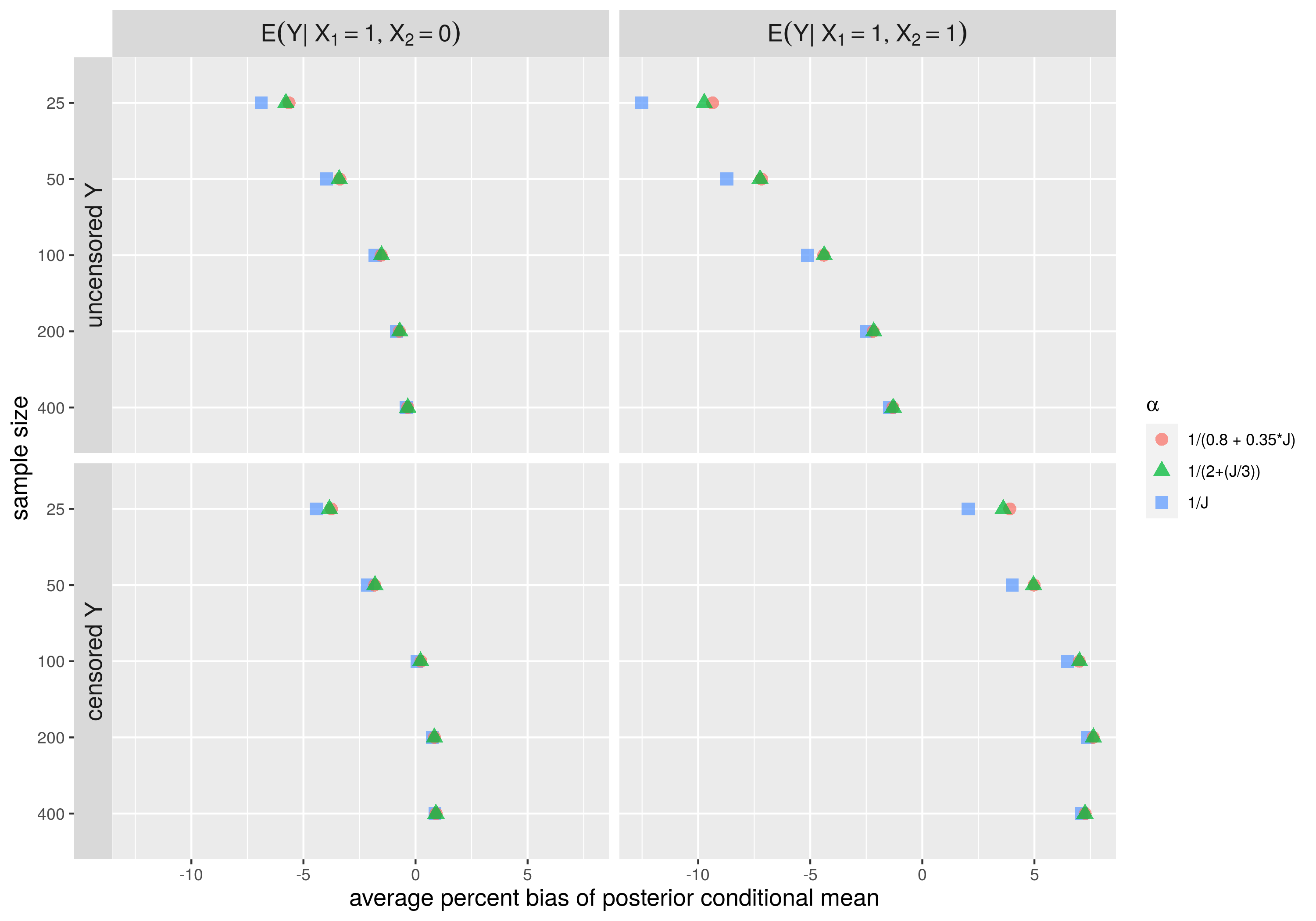} 

}

\caption{Percent bias in conditional mean for simulations using loglog link}\label{fig:simplt-mn-3}
\end{figure}

For scenario (3) the average percent bias for the uncensored outcome ranges from -12.5\% to -1.0\% with larger bias for the \(\boldsymbol{\alpha}=1/J\) prior and smaller \(n\) (Figure \ref{fig:simplt-mn-3}). As in the first two scenarios, the censored outcome (which replaced outcomes less than 0 with a value of \(y=0\)) showed a positive shift in average percent bias at \((X_1=1,X_2=1)\) although to a much smaller degree than scenario (1).

\hypertarget{conditional-median-and-quantiles}{%
\subsubsection{Conditional Median and Quantiles}\label{conditional-median-and-quantiles}}

\begin{figure}

{\centering \includegraphics[width=0.9\linewidth]{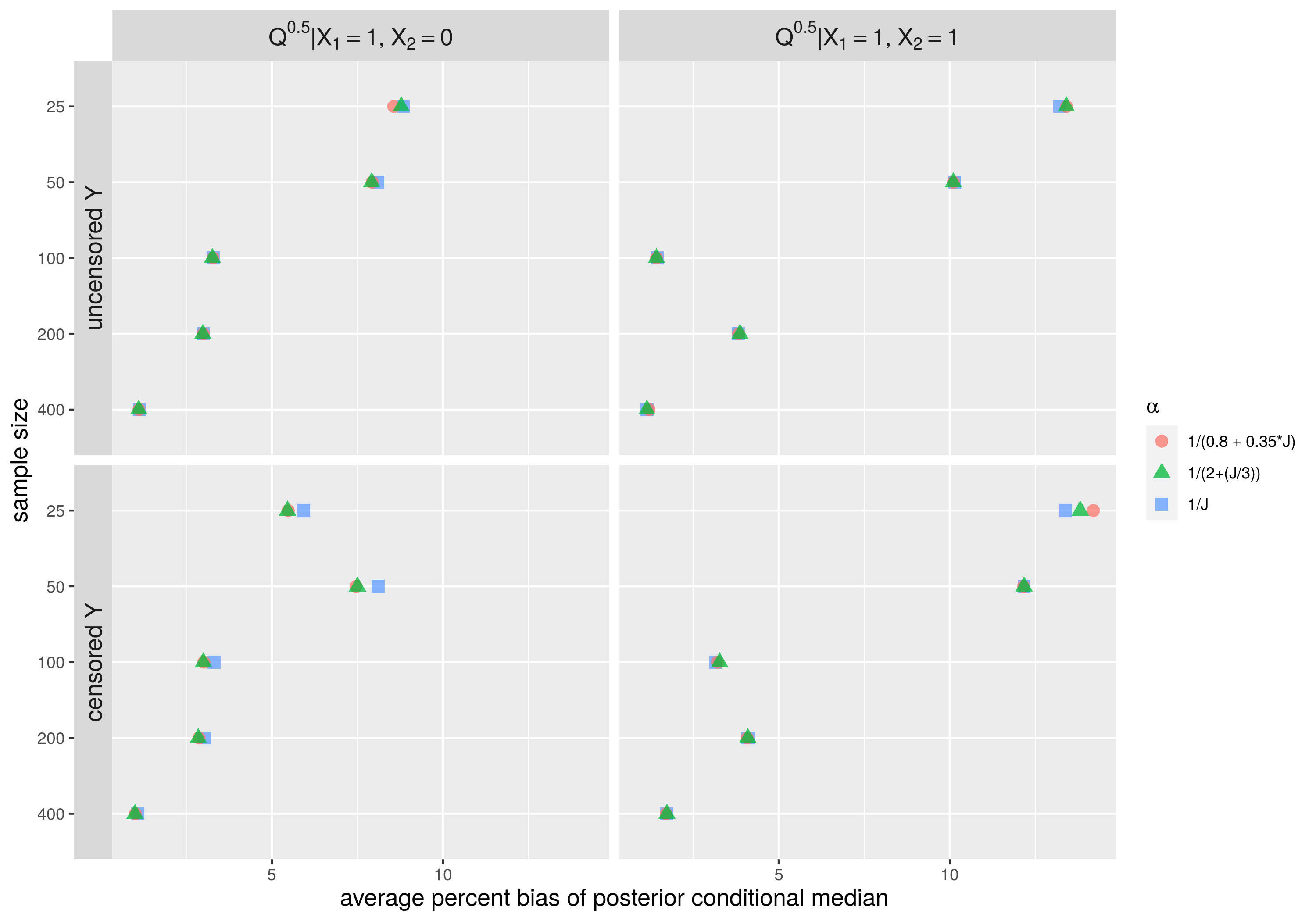} 

}

\caption{Percent bias in conditional median for simulations using probit link}\label{fig:simplt-med-1}
\end{figure}

\begin{figure}

{\centering \includegraphics[width=0.75\linewidth]{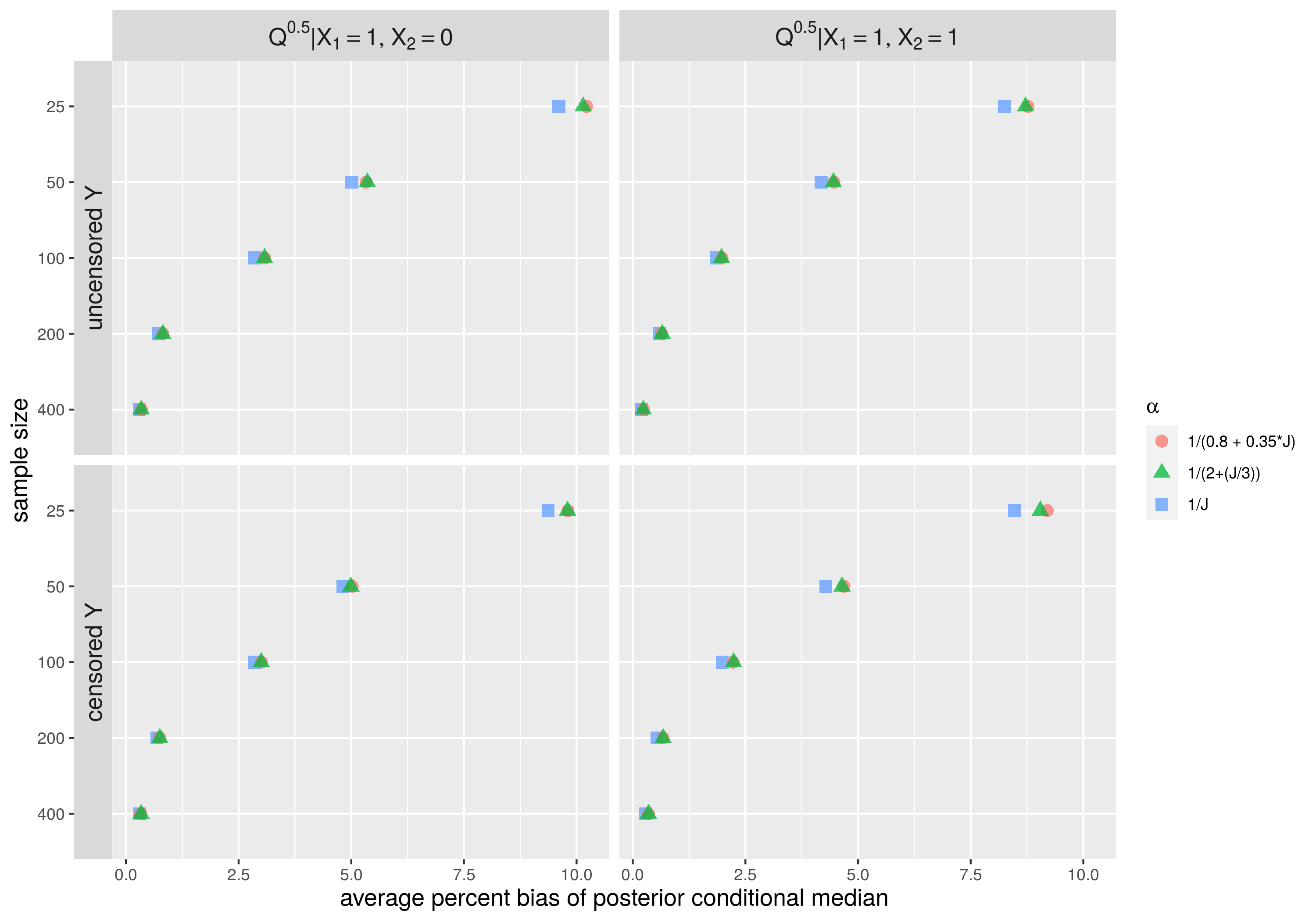} 

}

\caption{Percent bias in conditional median for simulations using logit link}\label{fig:simplt-med-2}
\end{figure}

\begin{figure}

{\centering \includegraphics[width=0.75\linewidth]{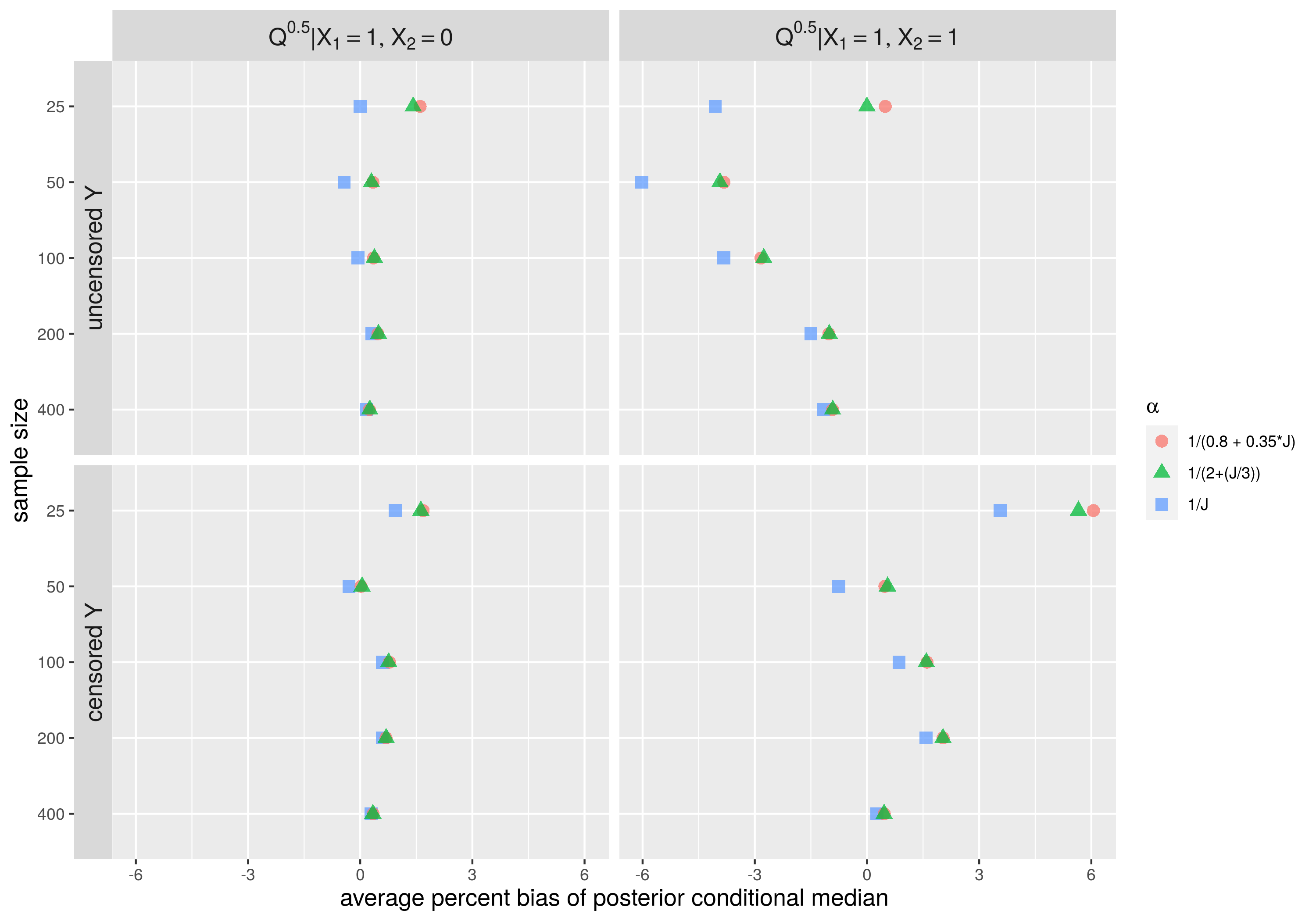} 

}

\caption{Percent bias in conditional median for simulations using loglog link}\label{fig:simplt-med-3}
\end{figure}

The simulation results for the conditional posterior median in scenario (1) are shown in Figure \ref{fig:simplt-med-1}. Across outcomes, the conditional median estimates had a positive average percent bias for both \((X_1=1,X_2=0)\) and \((X_1=1,X_2=1)\) with smaller bias for larger sample sizes where there was more information to estimate the center of the distribution. There were negligible differences in average percent bias of the median estimates for the three \(\boldsymbol{\alpha}\) concentration parameter priors. The pattern looked similar for scenario (2) (Supp. Figure \ref{fig:simplt-med-2}). For both outcomes under scenario (3), average percent bias in the conditional median estimate was smaller at \((X_1=1,X_2=0)\) than \((X_1=1,X_2=1)\). There were only small differences between the three \(\boldsymbol{\alpha}\) concentration parameters except with the smaller sample sizes (Supp. Figure \ref{fig:simplt-med-3}).

\begin{figure}

{\centering \includegraphics[width=0.9\linewidth]{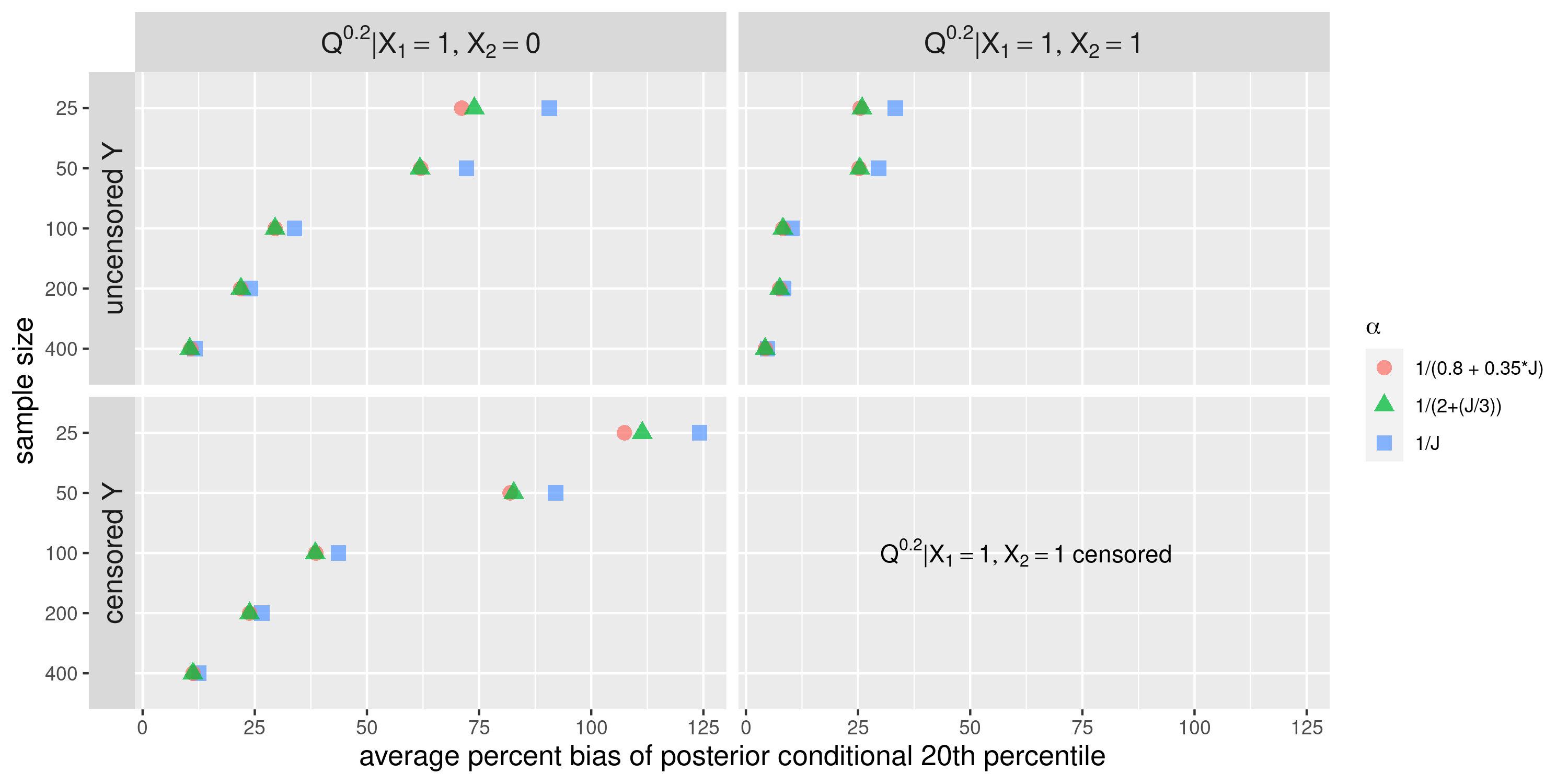} 

}

\caption{Percent bias in conditional 20th percentile for simulations using probit link}\label{fig:simplt-q20-a}
\end{figure}

\begin{figure}

{\centering \includegraphics[width=0.75\linewidth]{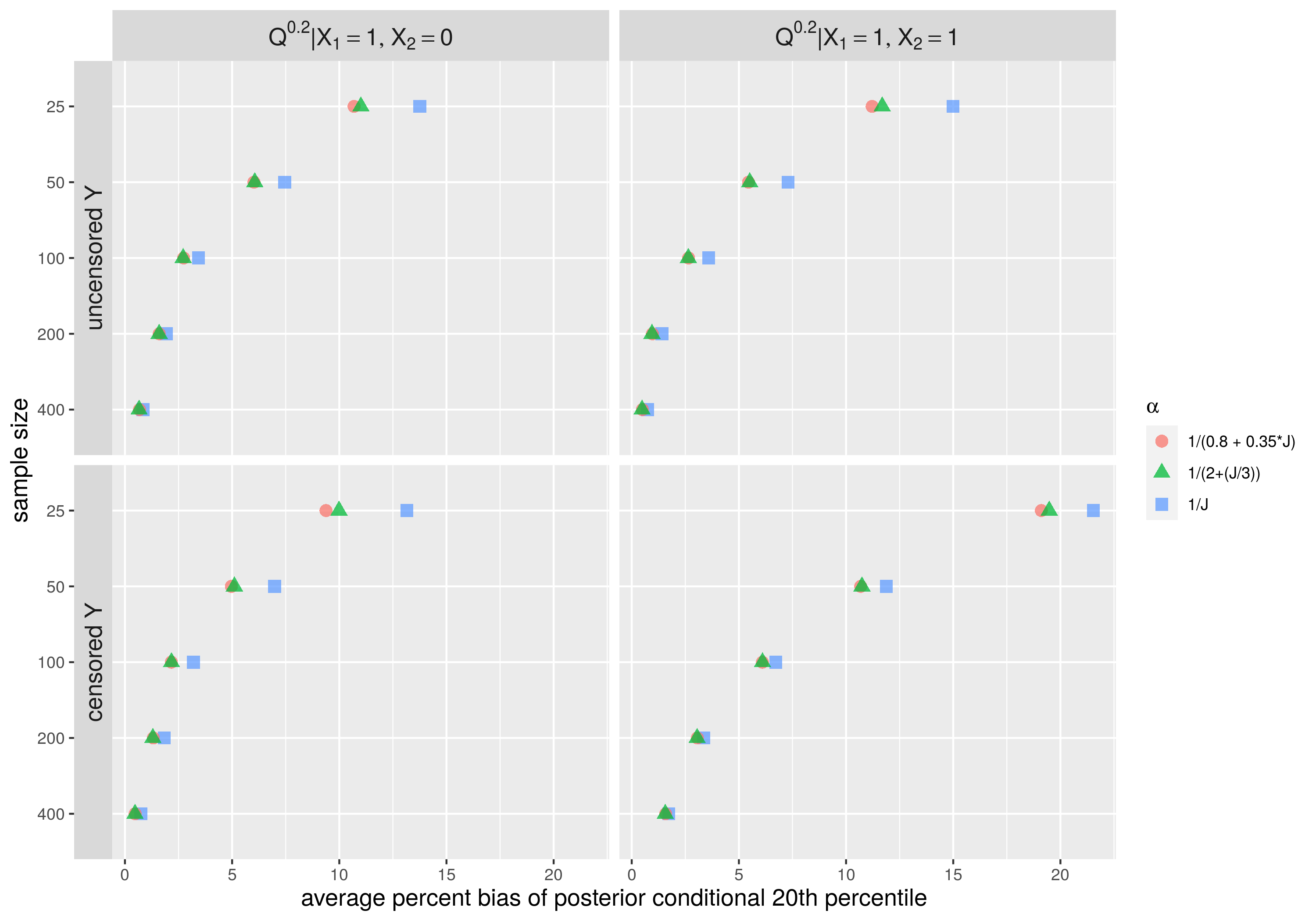} 

}

\caption{Bias in conditional 20th percentile for simulations using logit link}\label{fig:simplt-q20-b}
\end{figure}

\begin{figure}

{\centering \includegraphics[width=0.75\linewidth]{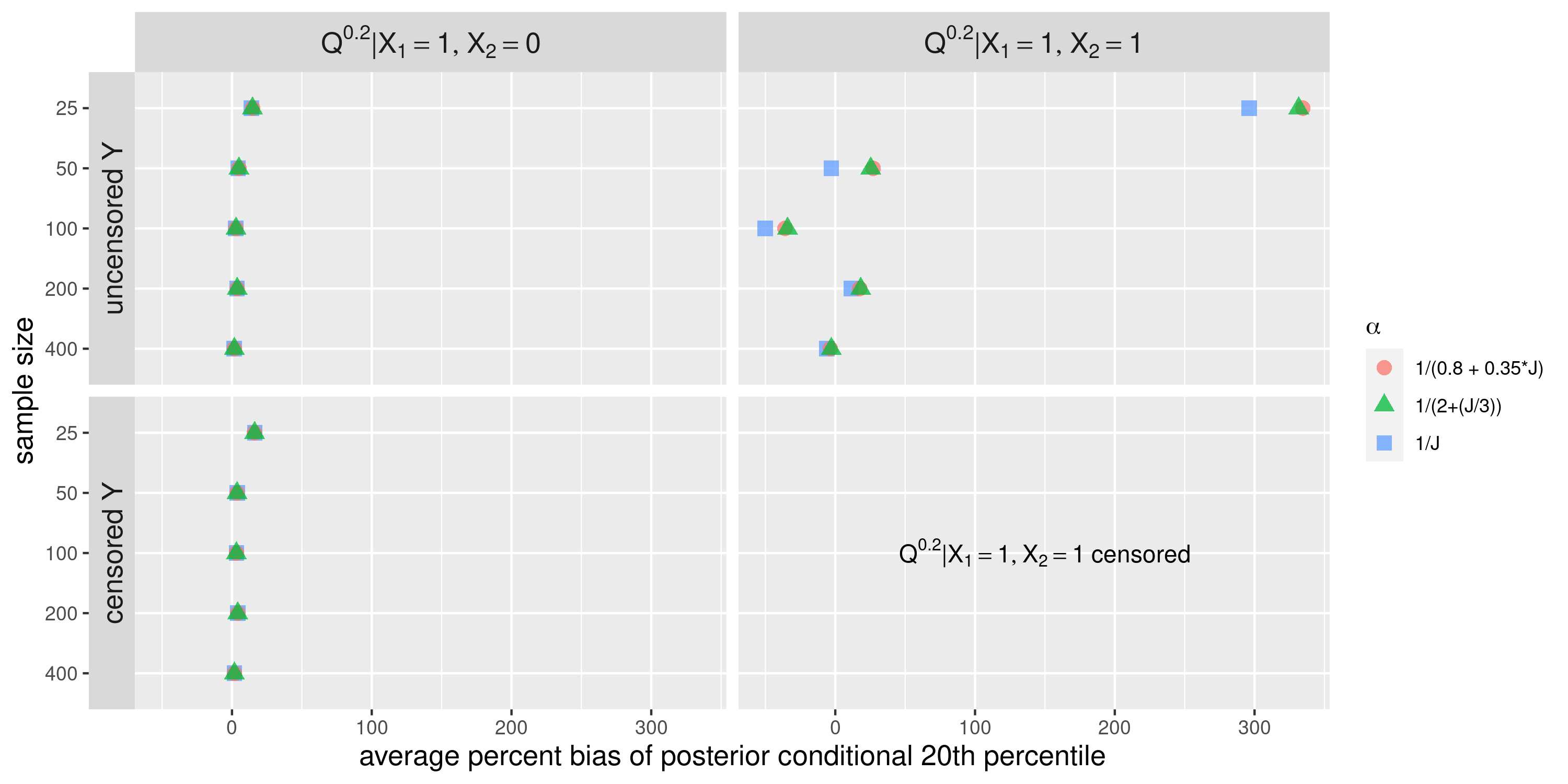} 

}

\caption{Bias in conditional 20th percentile for simulations using loglog link}\label{fig:simplt-q20-c}
\end{figure}

Figure \ref{fig:simplt-q20-a} presents the results for the posterior conditional 20th percentile in scenario (1). The uncensored outcome estimates were quite biased (between 25\% and 90\%) for the smaller sample sizes. The magnitude of the bias varied based on the values of the conditioning variables, \(X_1\) and \(X_2\), with larger bias when the conditional distribution was further from \(\beta_1 = \beta_2 = 0\). The estimates of the conditional 20th percentile for the censored outcome in scenario (1) were similar to the uncensored outcome. When \((X_1=1,X_2=1)\) the true conditional \(Q^{0.2}\) falls below the censoring threshold and does not have a specific numeric value. In this case percent bias could not be computed. For scenario (2) the estimates of the posterior conditional 20th percentile with the uncensored outcome were again positively biased for the smaller sample sizes with more bias for the \(\boldsymbol{\alpha}=1/J\) concentration prior (Supp. Figure \ref{fig:simplt-q20-b}). Under scenario (3) the uncensored outcome estimates of the conditional 20th percentile had reasonably small average percent bias for all the priors and sample sizes except at \((X_1=1,X_2=1)\) when \(n=25\). Similar to scenario (1), the censored outcome estimates showed small average percent bias at \((X_1=1,X_2=0)\), but the true conditional 20th percentile fell below the censoring threshold for \((X_1=1,X_2=1)\) precluding calculation of percent bias (Supp. Figure \ref{fig:simplt-q20-c}).

\hypertarget{computation-time}{%
\subsubsection{Computation time}\label{computation-time}}

\begin{figure}

{\centering \includegraphics[width=0.95\linewidth]{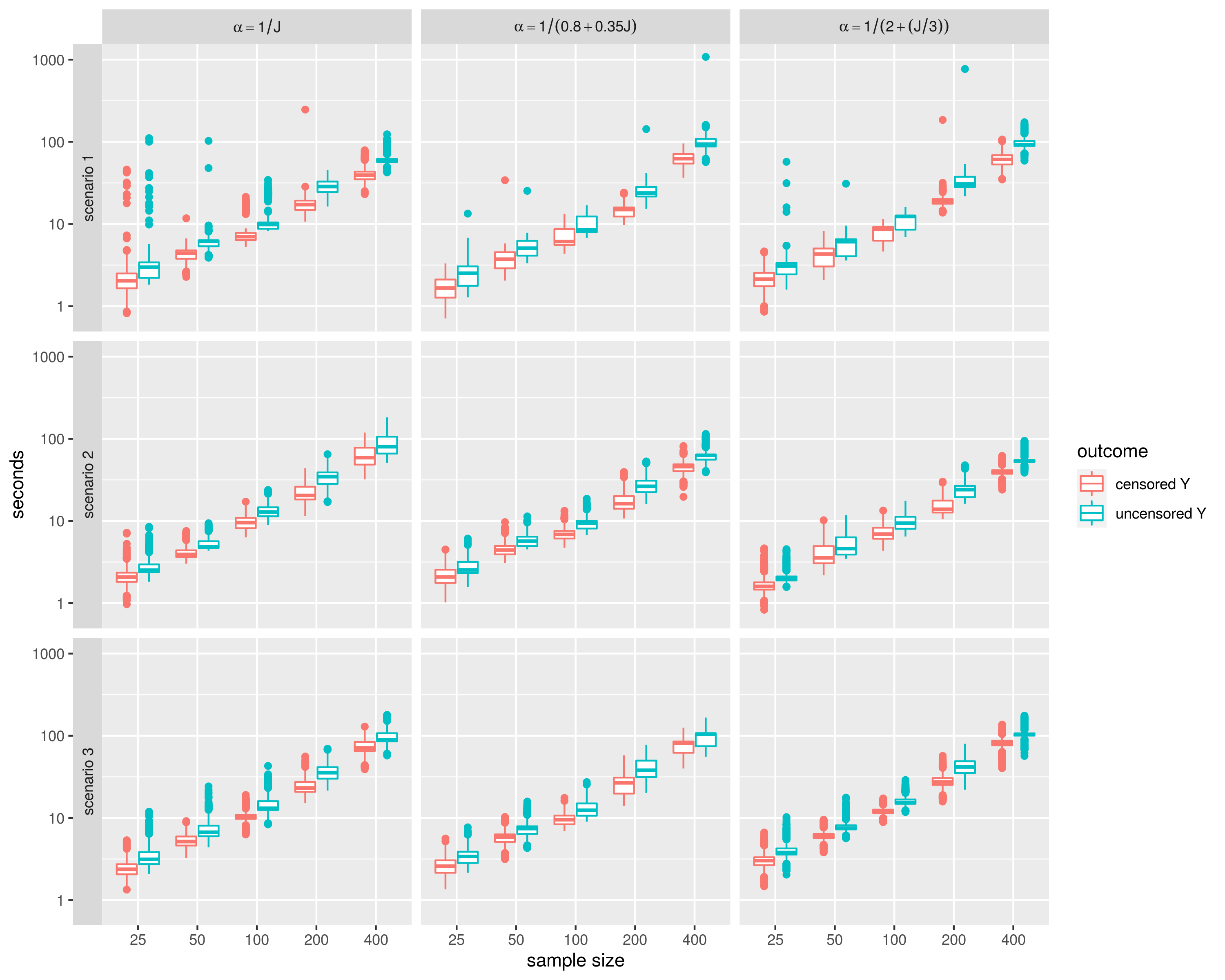} 

}

\caption{Per chain MCMC sampling time for three simulation scenarios. Each boxplot shows the sampling times required to produce 4000 posterior draws under the specified model/prior/sample size combination for 1000 simulation datasets}\label{fig:simplt-samptime}
\end{figure}

Simulations were performed using \texttt{R} version 3.6.0 (2019-04-26) and \texttt{rstan} (Version 2.19.2) on a high-performance computing cluster running under CentOS Linux 7 (Core) with 1.90GHz or 2.40GHz Intel Xeon CPUs and up to 3 GB of memory per compute node. MCMC sampling time for the three scenarios is shown in Figure \ref{fig:simplt-samptime}. Per chain sampling time increased approximately exponentially with sample size and was similar across scenarios and priors.

\hypertarget{case-study}{%
\section{4. Case Study}\label{case-study}}

\hypertarget{background-and-methods}{%
\subsection{Background and Methods}\label{background-and-methods}}

The data for the case study were collected from 216 HIV-positive adults on antiretroviral therapy in two cohort studies (Vanderbilt Lipoatrophy and Neuropathy Cohort (LiNC), n=147; Adiposity and Immune Activation Cohort (AIAC), n=69). Further details on the study design and cohorts are provided in Koethe et al. {[}\protect\hyperlink{ref-koethe_serum_2012}{33},\protect\hyperlink{ref-koethe_metabolic_2015}{34}{]}. Because people living with HIV have increased risk of diabetes and cardiovascular disease, the aim of the analysis was to estimate the association between body mass index (BMI) and several inflammation biomarkers in this population, adjusting for additional covariates: age, sex, race, smoking status, study location and CD4 cell count.

We examine the biomarkers Interleukin 6 (IL-6) and Interleukin 1 beta (IL-1-\(\beta\)); both are right-skewed with 3\% and 39\% of values censored below the lower limit of detection, respectively. Censored values are set to 0. To account for skewness and censoring we fit Bayesian CPMs using logit, probit, and loglog link functions, noninformative \(\beta\) priors and a concentration parameter of either \(\boldsymbol{\alpha}=1/J\) or \(\boldsymbol{\alpha}=1/(0.8+0.35J)\) for the Dirichlet prior to estimate the association between BMI and the conditional mean, median, and 90th percentile of each biomarker.

We evaluate convergence using \(\hat{R}\) scale reduction factor {[}\protect\hyperlink{ref-gelman_bayesian_2014}{29}{]} and traceplots of MCMC draws. Model comparison is performed using the difference in expected log predictive density (ELPD) calculated using leave-one-out cross-validation {[}\protect\hyperlink{ref-vehtari_practical_2017}{35}{]}. Model fit is assessed with graphical checks of the posterior predictive distribution and posterior predictive p-values {[}\protect\hyperlink{ref-gelman_bayesian_2014}{29},\protect\hyperlink{ref-stern_bayesian_2005}{36}{]}.

\hypertarget{results-1}{%
\subsection{Results}\label{results-1}}

For each of the two biomarker outcomes six model specifications were fit: probit, logit, or loglog link with \(\boldsymbol{\alpha}=1/(0.8+0.35J)\) or \(\boldsymbol{\alpha}=1/J\). Each model sampled from 2 chains with 2000 warmup and 4000 total iterations to produce 4000 posterior sample draws for each parameter. For all models, traceplots showed no issues with mixing or stationarity; further, all \(\hat{R}\) potential scale reduction values were \(<1.01\) indicating likely convergence. Table \ref{tab:elpdtab} shows the difference in ELPD for the IL-6 and IL-1-\(\beta\) biomarker models. Based on the difference in ELPD, the CPM with loglog link and \(\boldsymbol{\alpha}=1/(0.8+0.35J)\) was used for the both outcomes, however there is little difference in ELPD along the top several models.

\begin{table}

\caption{\label{tab:elpdtab}Difference in expected log pointwise predictive density for IL-6 models and IL-1-$\beta$ models}
\centering
\begin{tabular}[t]{lrr}
\toprule
Model & ELPD diff. & SE diff.\\
\midrule
\addlinespace[0.3em]
\multicolumn{3}{l}{\textbf{IL-6}}\\
\hspace{1em}loglog link, $\boldsymbol{\alpha}=1/(0.8+0.35J)$ & 0.00 & \vphantom{1} 0.00\\
\hspace{1em}logit link, $\boldsymbol{\alpha}=1/J$ & -1.94 & 6.65\\
\hspace{1em}probit link, $\boldsymbol{\alpha}=1/(0.8+0.35J)$ & -1.99 & 6.64\\
\hspace{1em}logit link, $\boldsymbol{\alpha}=1/(0.8+0.35J)$ & -4.96 & 6.68\\
\hspace{1em}loglog link, $\boldsymbol{\alpha}=1/J$ & -7.16 & 5.38\\
\hspace{1em}probit link, $\boldsymbol{\alpha}=1/J$ & -7.40 & 6.87\\
\addlinespace[0.3em]
\multicolumn{3}{l}{\textbf{IL-1-$\beta$}}\\
\hspace{1em}loglog link, $\boldsymbol{\alpha}=1/(0.8+0.35J)$ & 0.00 & 0.00\\
\hspace{1em}probit link, $\boldsymbol{\alpha}=1/(0.8+0.35J)$ & -2.82 & 4.80\\
\hspace{1em}logit link, $\boldsymbol{\alpha}=1/(0.8+0.35J)$ & -6.45 & 5.26\\
\hspace{1em}probit link, $\boldsymbol{\alpha}=1/J$ & -7.25 & 4.85\\
\hspace{1em}loglog link, $\boldsymbol{\alpha}=1/J$ & -10.81 & 4.03\\
\hspace{1em}logit link, $\boldsymbol{\alpha}=1/J$ & -11.70 & 4.61\\
\bottomrule
\end{tabular}
\end{table}

\hypertarget{il-6-biomarker}{%
\subsubsection{IL-6 biomarker}\label{il-6-biomarker}}

\begin{figure}

{\centering \includegraphics[width=0.75\linewidth]{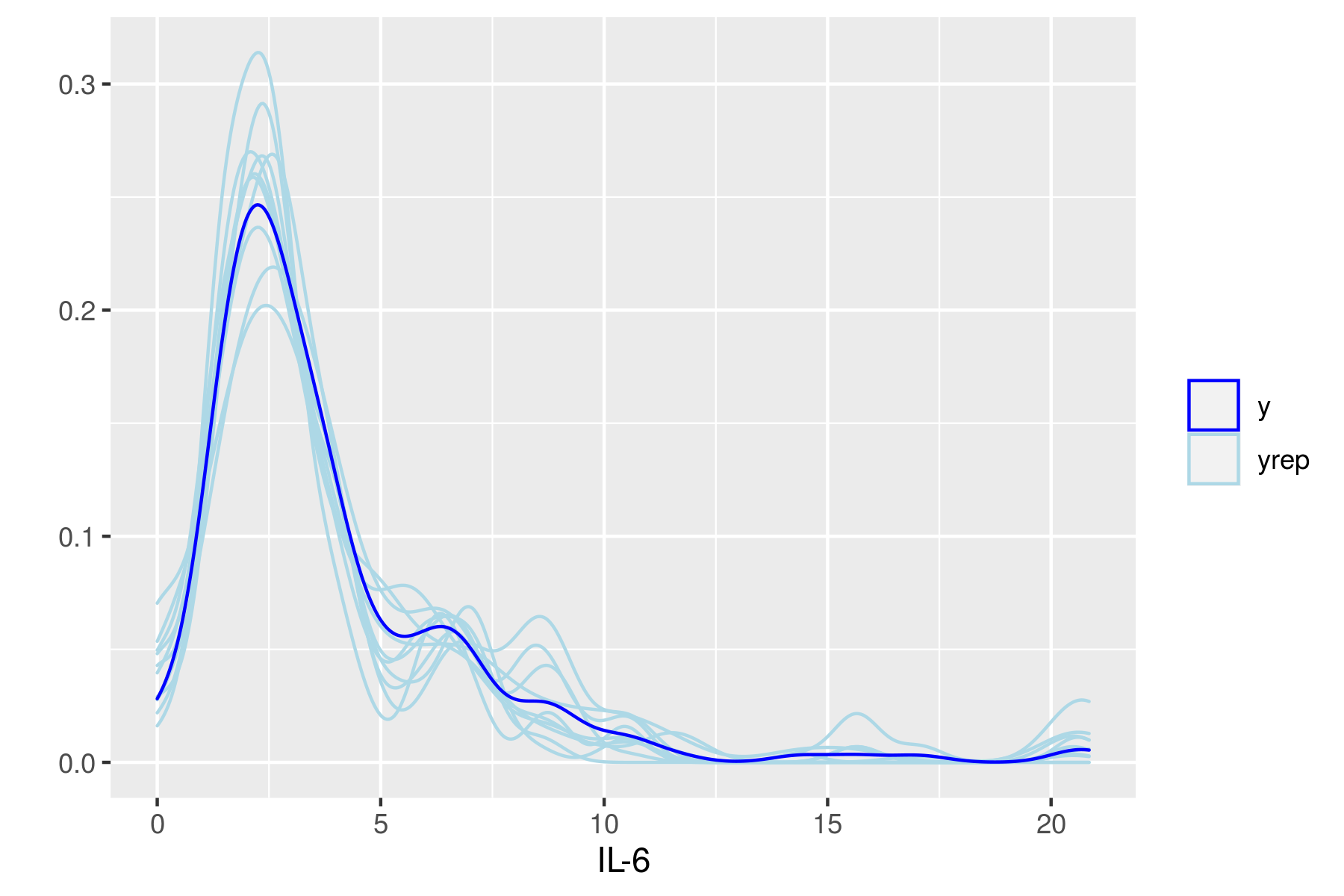} 

}

\caption{Observed outcome ($y$) and 10 posterior predictive distribution draws ($y_{rep}$) for IL-6 model}\label{fig:il6-postpred}
\end{figure}

A graphical check of 10 draws from the posterior predictive distribution compared to the observed IL-6 distribution (Figure \ref{fig:il6-postpred}) did not indicate any serious model misfit. In addition, there were no major discrepancies between the model and data based on the posterior predictive p-values for the test quantities variance, skewness, and proportion of observations censored below the lower limit of detection (Table \ref{tab:ppptab1}) so the CPM was able to reproduce these aspects of the observed data fairly well.

\begin{table}[!h]

\caption{\label{tab:ppptab1}Posterior predictive p-values for IL-6 model}
\centering
\begin{tabular}[t]{lr}
\toprule
Test quantity & Posterior predictive p-value\\
\midrule
variance & 0.43\\
skewness & 0.34\\
proportion censored & 0.53\\
\bottomrule
\end{tabular}
\end{table}

\begin{figure}

{\centering \includegraphics[width=0.49\linewidth]{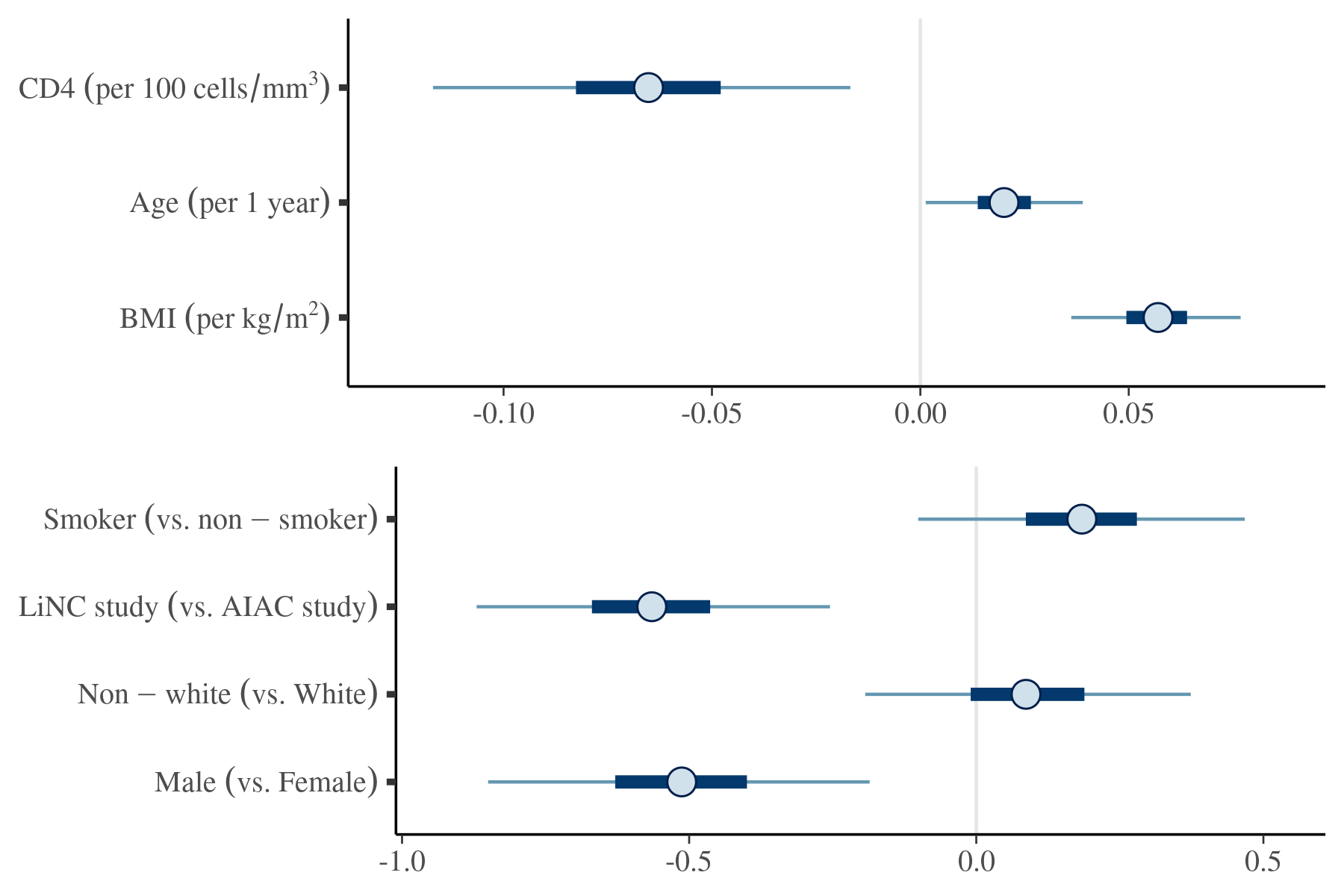} \includegraphics[width=0.49\linewidth]{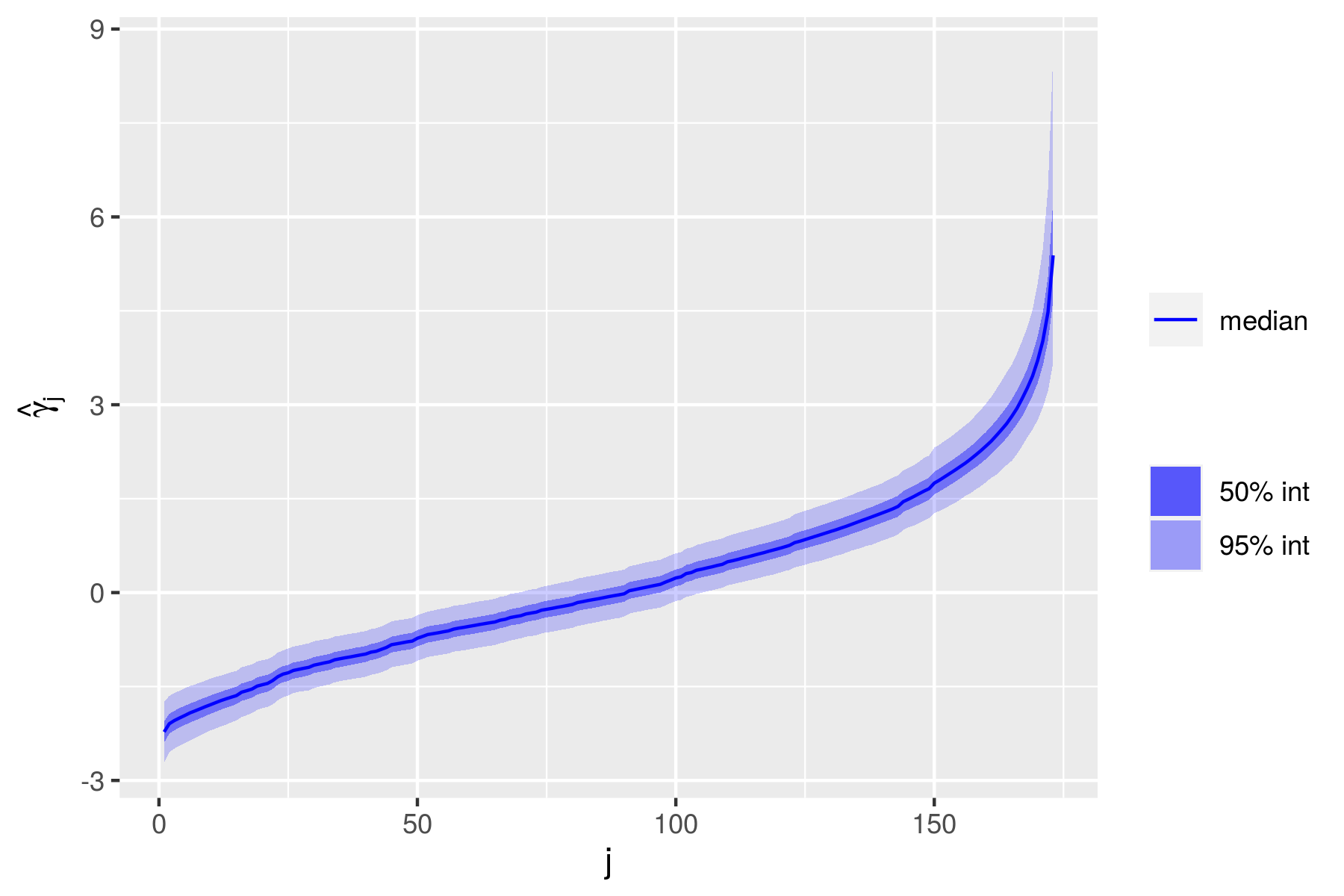} 

}

\caption{(a) Posterior median $\boldsymbol{\beta}$ estimates and (b) posterior median $\boldsymbol{\gamma}$ estimates with 50\% and 95\% credible intervals for IL-6 model}\label{fig:il6-par00}
\end{figure}

\begin{figure}

{\centering \includegraphics[width=0.75\linewidth]{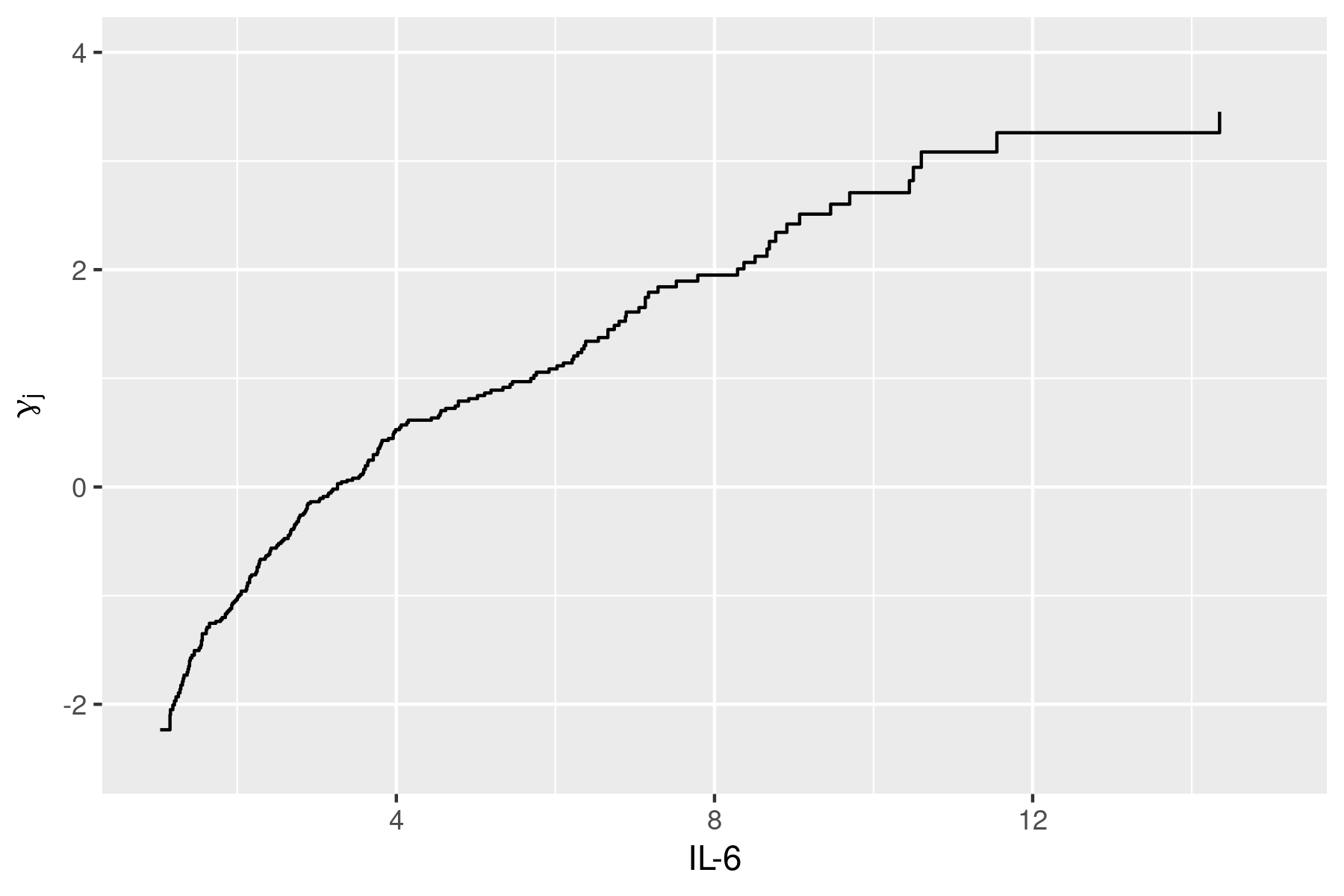} 

}

\caption{Estimated transformation for IL-6 model}\label{fig:il-6-trans-1}
\end{figure}

The median posterior estimates of the covariate parameters along with 50\% and 95\% credible intervals for the IL-6 model are shown in Figure \ref{fig:il6-par00}a. Age and BMI were positively associated with increased IL-6, while CD4 count, male gender, and the Lipoatrophy and Neuropathy cohort were negatively associated with IL-6. The relationship between IL-6 and smoking and nonwhite race was more equivocal. Figure \ref{fig:il6-par00}b shows the posterior median \(\boldsymbol{\gamma}\) estimates along with the 50\% and 95\% credible intervals. Plotting the \(\boldsymbol{\gamma}\) estimates against the observed IL-6 values (Figure \ref{fig:il-6-trans-1}) gives the estimated transformation, \(\hat{H}\).

\begin{figure}

{\centering \includegraphics[width=0.75\linewidth]{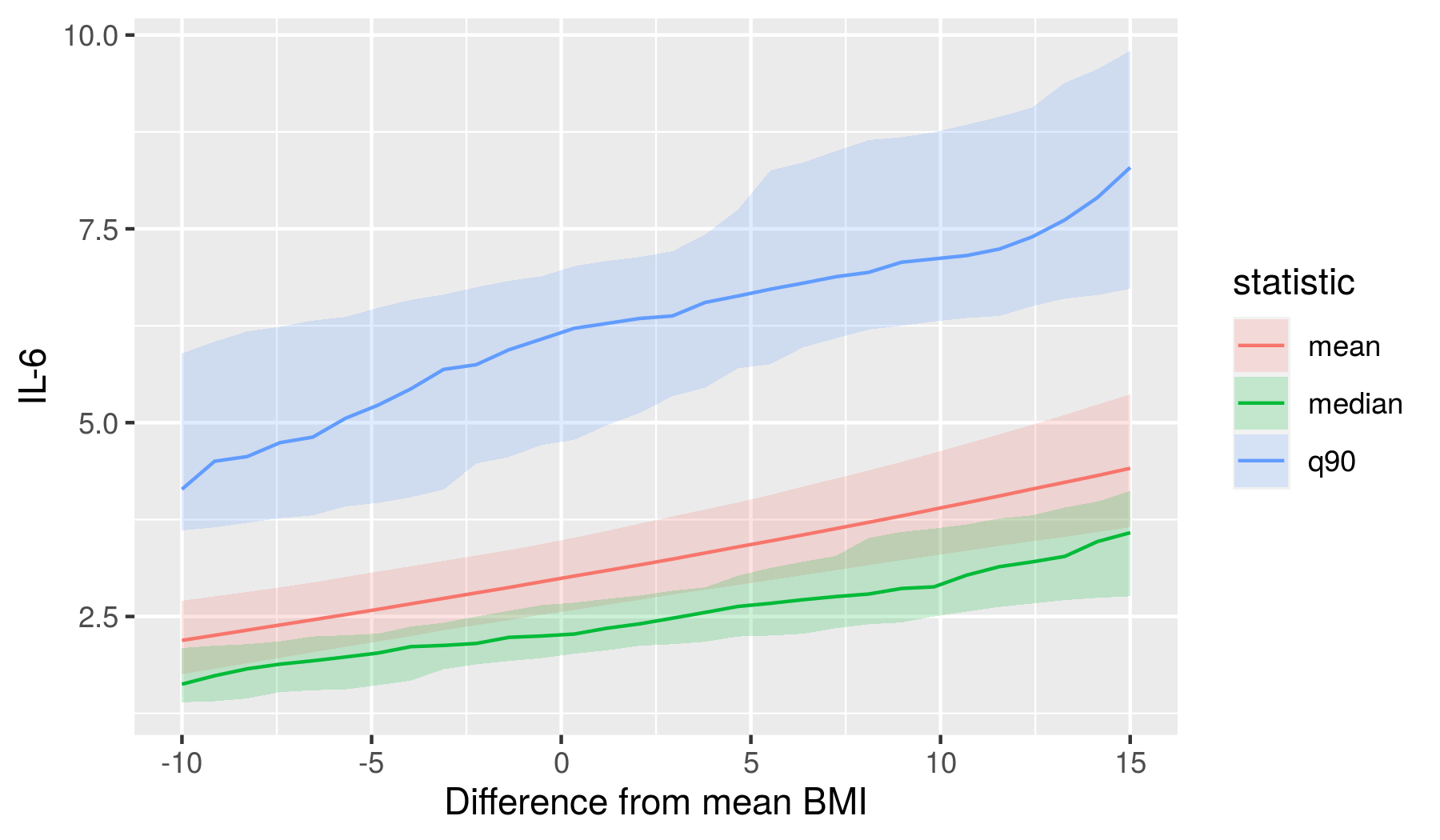} 

}

\caption{Difference from mean BMI vs. IL-6 mean, median, and 90th percentile for a white, male, nonsmoker with average age and CD4 count in the Lipoatrophy and Neuropathy cohort}\label{fig:il6-bmi}
\end{figure}

The estimated relationship between BMI and the posterior conditional mean (using 0 for censored values), median, and 90th percentile of IL-6 (for a white, male, nonsmoker with average age and CD4 count in the LiNC study) is shown in Figure \ref{fig:il6-bmi} along with 95\% credible intervals. Higher BMI was associated with higher IL-6.

\hypertarget{il-1-beta-biomarker}{%
\subsubsection{\texorpdfstring{IL-1-\(\beta\) biomarker}{IL-1-\textbackslash beta biomarker}}\label{il-1-beta-biomarker}}

\begin{figure}

{\centering \includegraphics[width=0.75\linewidth]{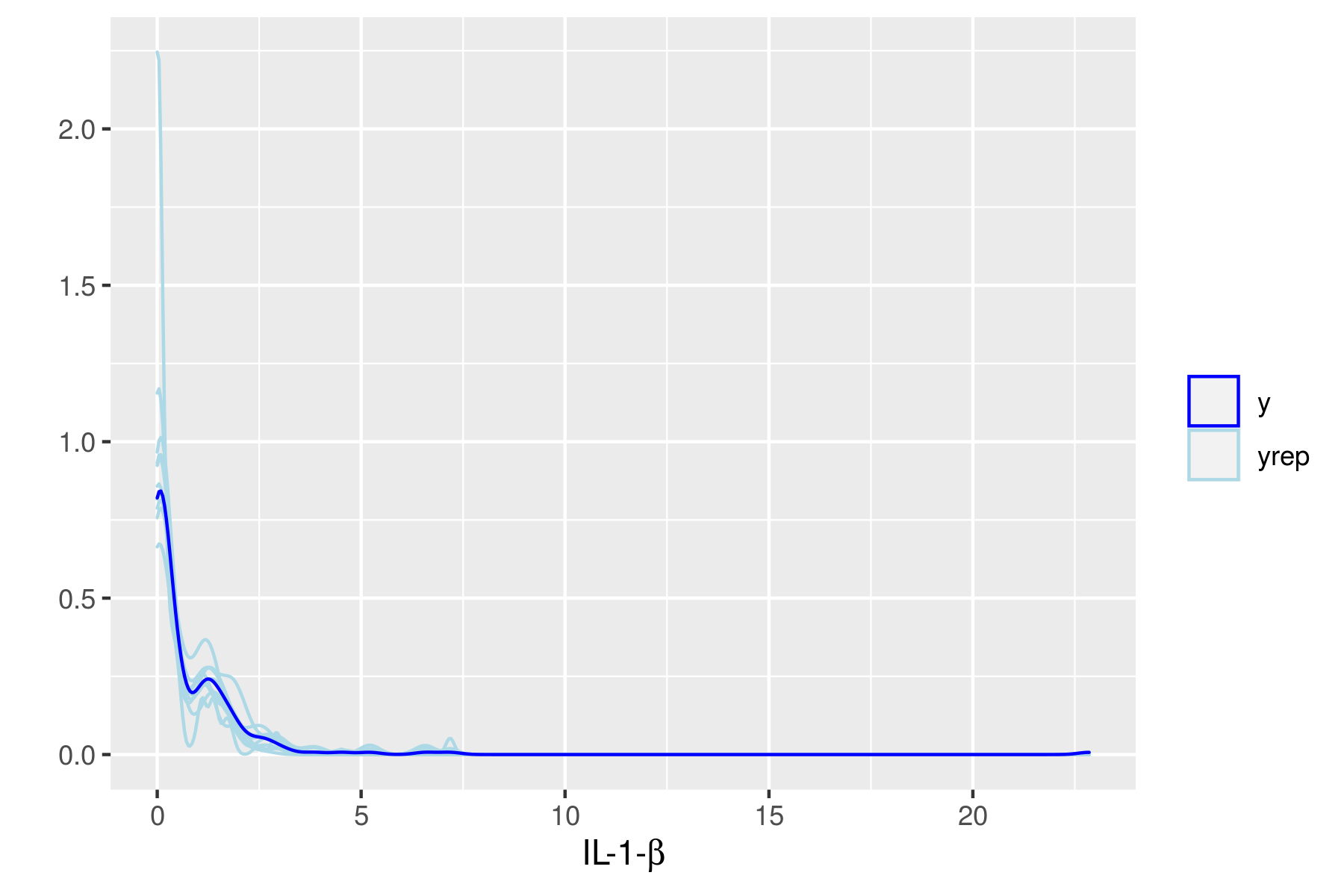} 

}

\caption{Observed outcome ($y$) and 10 posterior predictive distribution draws ($y_{rep}$) for IL-1-$\beta$ model}\label{fig:il-1-beta-postpred}
\end{figure}

As with the IL-6 biomarker, comparing the observed IL-1-\(\beta\) distribution to draws from the posterior predictive distribution (Figure \ref{fig:il-1-beta-postpred}) did not reveal any serious model misfit. The posterior predictive p-values for variance, skewness, and proportion of observations below the lower limit of detection are shown in Table \ref{tab:ppptab2}. There was no indication of serious discrepancy between the model and data for variance and proportion of censored observations although the posterior predictive p-value for skewness was more extreme indicating a moderate degree of misfit. This seems reasonable given the high level of right-skewness for IL-1-\(\beta\).

\begin{table}[!h]

\caption{\label{tab:ppptab2}Posterior predictive p-values for IL-1-$\beta$ model}
\centering
\begin{tabular}[t]{lr}
\toprule
Test quantity & Posterior predictive p-value\\
\midrule
variance & 0.38\\
skewness & 0.15\\
proportion censored & 0.63\\
\bottomrule
\end{tabular}
\end{table}

\begin{figure}

{\centering \includegraphics[width=0.49\linewidth]{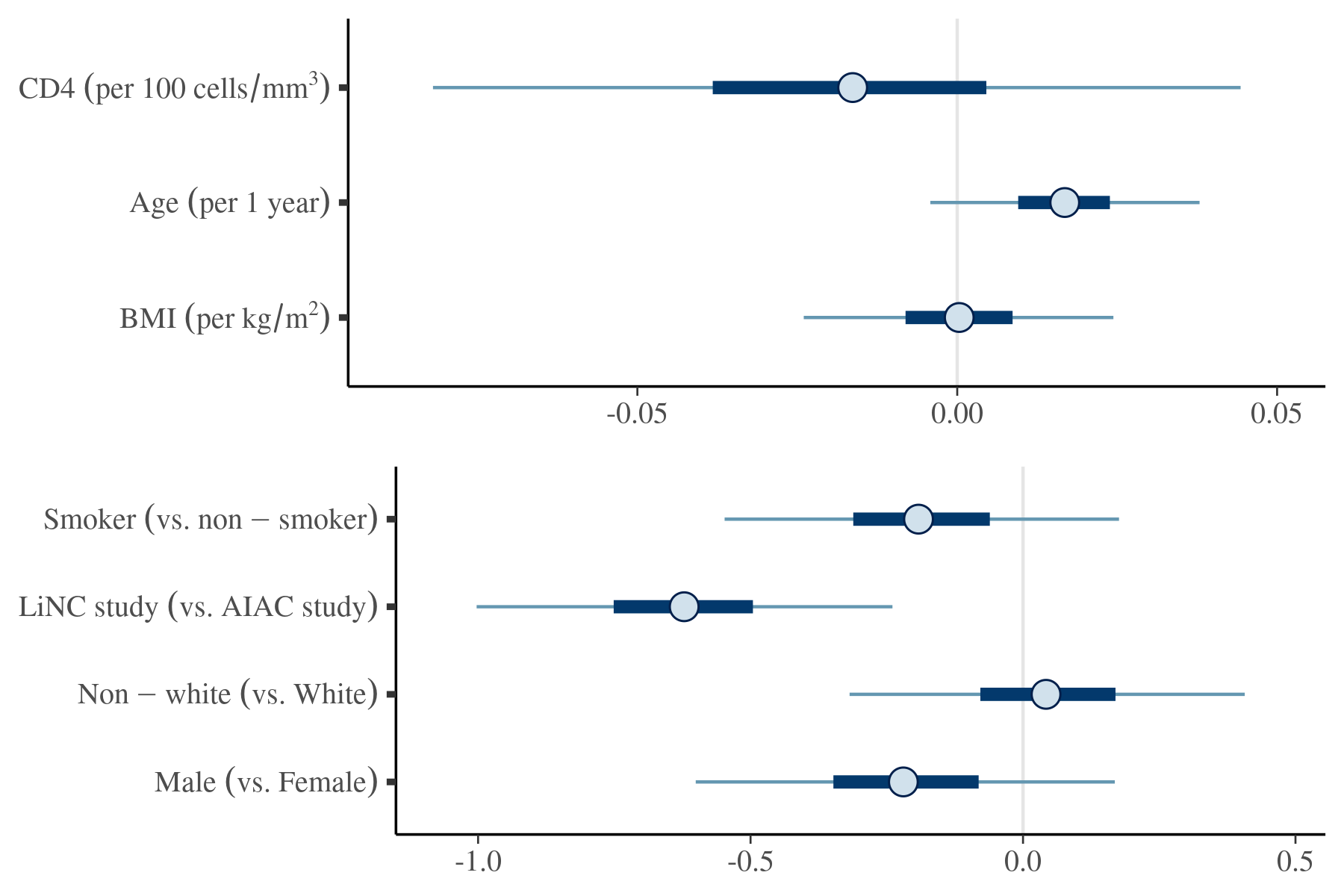} \includegraphics[width=0.49\linewidth]{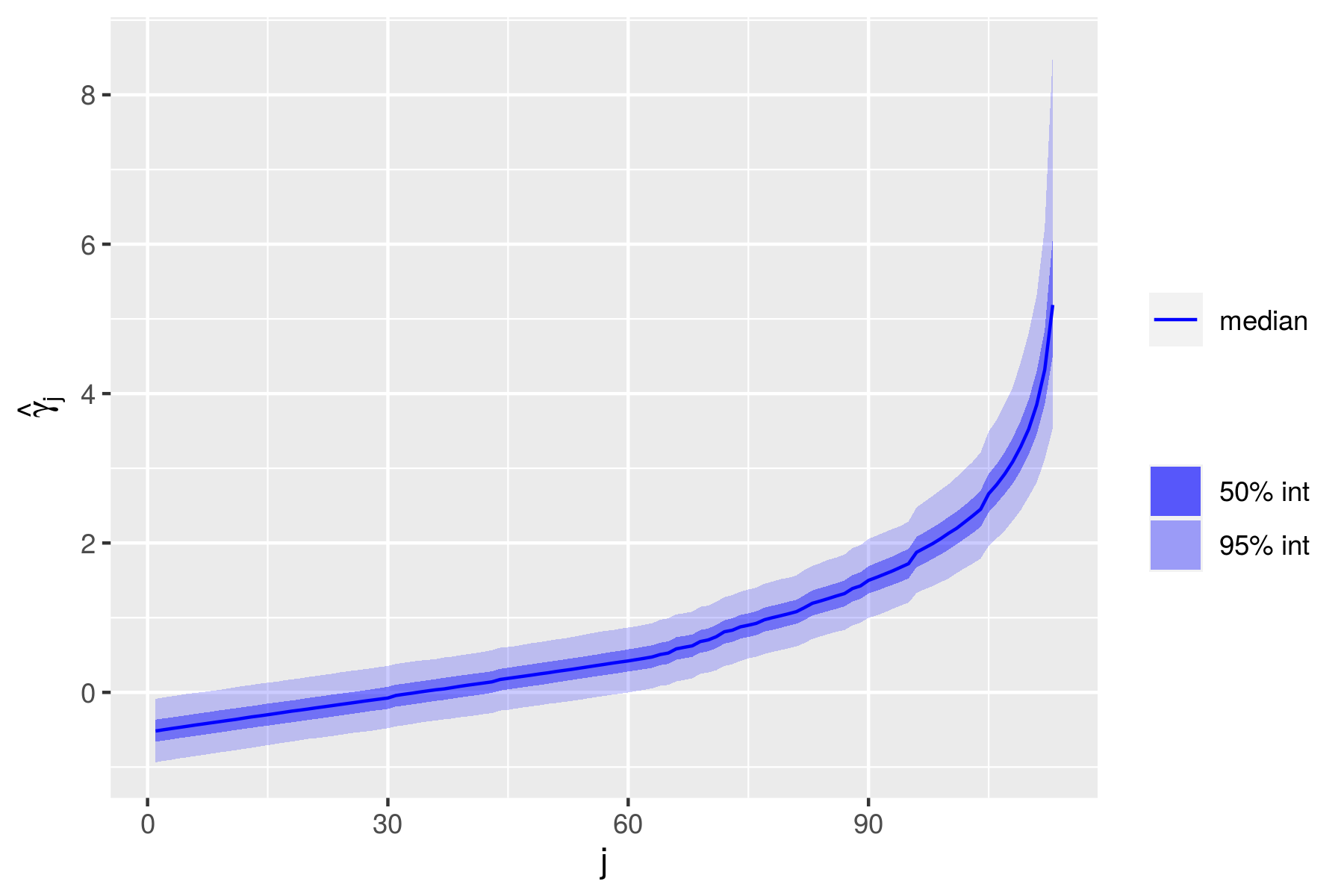} 

}

\caption{(a) Posterior median $\boldsymbol{\beta}$ estimates and (b) posterior median $\boldsymbol{\gamma}$ estimates with 50\% and 95\% credible intervals for IL-1-$\beta$ model}\label{fig:il-1-beta-par00}
\end{figure}

\begin{figure}

{\centering \includegraphics[width=0.75\linewidth]{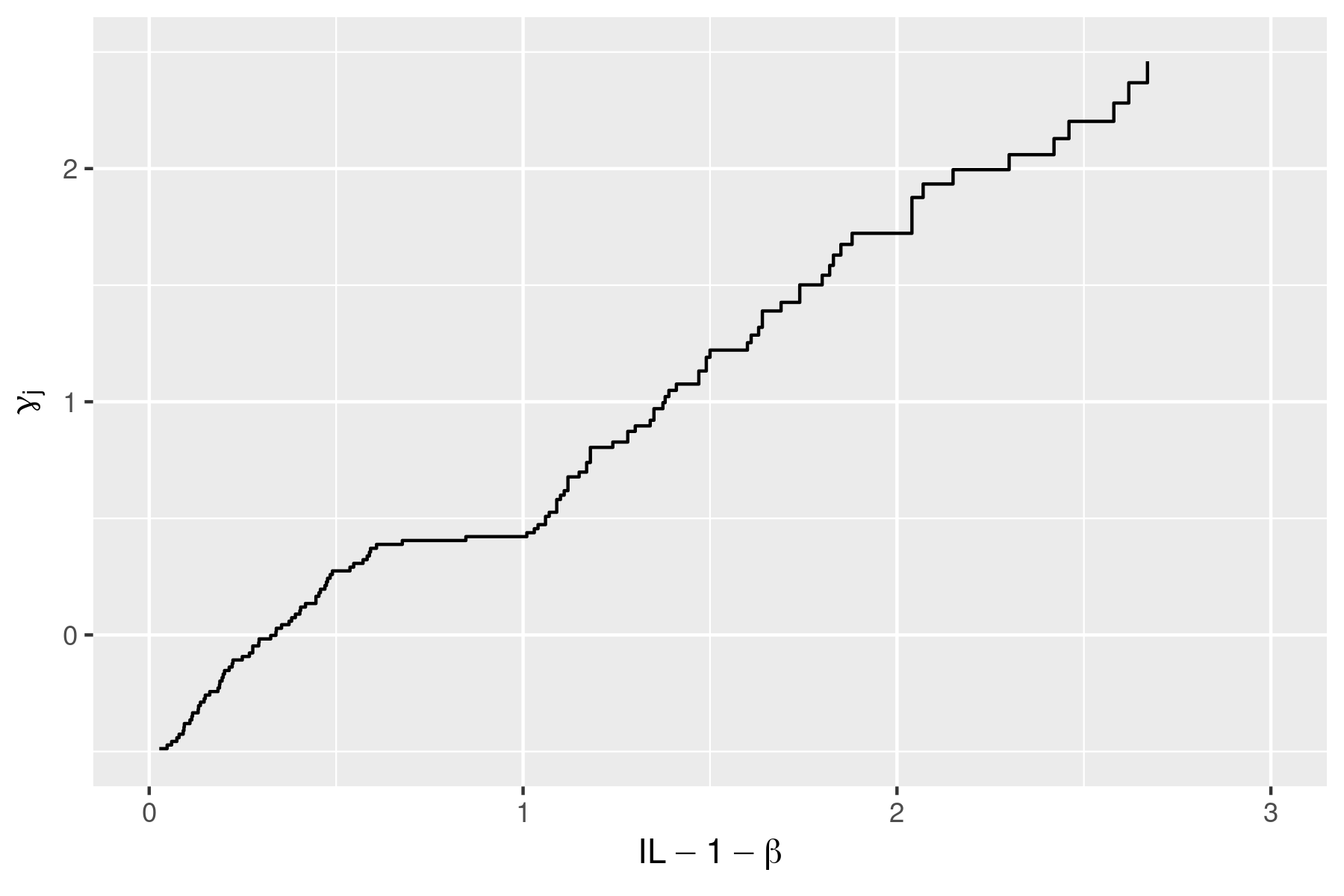} 

}

\caption{Estimated transformation for IL-1-$\beta$ model}\label{fig:il-1-beta-trans}
\end{figure}

The median posterior estimates of the covariate parameters along with 50\% and 95\% credible intervals for the IL-1-\(\beta\) model are shown in \ref{fig:il-1-beta-par00}a. In contrast to IL-6, there was weak association between all covariates (except study cohort) and IL-1-\(\beta\) level. Figure \ref{fig:il-1-beta-par00}b shows the posterior median \(\boldsymbol{\gamma}\) estimates along with the 50\% and 95\% credible intervals. Plotting the \(\boldsymbol{\gamma}\) estimates against the observed IL-1-\(\beta\) values gives the estimated transformation, \(\hat{H}\) (Figure \ref{fig:il-1-beta-trans}).

\begin{figure}

{\centering \includegraphics[width=0.75\linewidth]{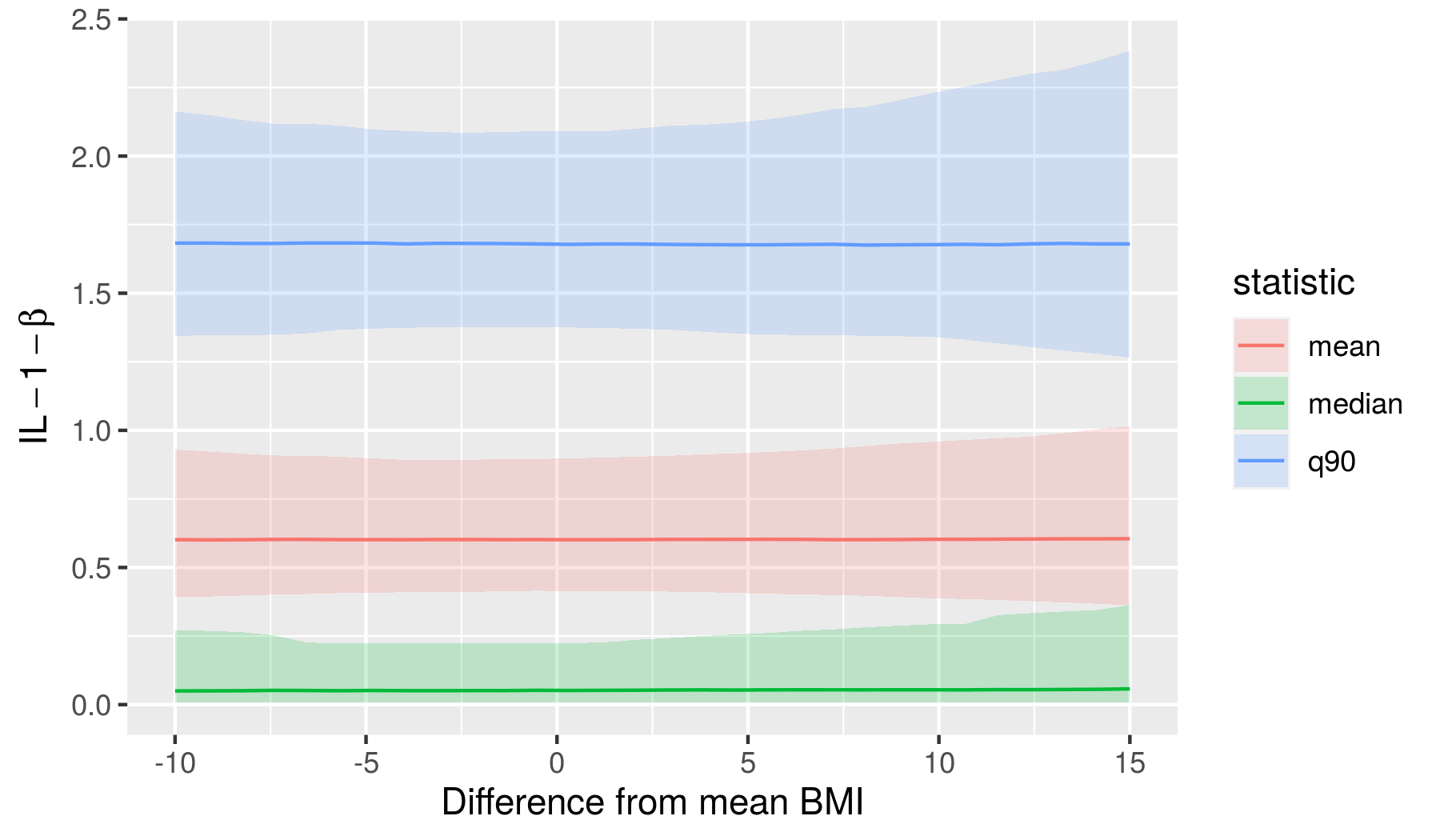} 

}

\caption{Difference from mean BMI vs. IL-1-$\beta$ mean, median, 90th percentile for a white, male, nonsmoker with average age and CD4 count in the Lipoatrophy and Neuropathy cohort}\label{fig:il-1-beta-bmi}
\end{figure}

Figure \ref{fig:il-1-beta-bmi} displays the estimated relationship between BMI and the posterior conditional mean (plugging in 0 for censored values), median, and 90th percentile of IL-1-\(\beta\) (for a white, male, nonsmoker with average age and CD4 count in the Lipoatrophy and Neuropathy cohort) along with 95\% credible intervals. The plot confirms little association between BMI and IL-1-\(\beta\).

\hypertarget{discussion}{%
\section{5. Discussion}\label{discussion}}

Although Bayesian CPM models have been frequently applied to ordinal data when the number of outcome categories is much smaller than the sample size, the extension to continuous or mixed outcomes where the number of categories is close or equal to the sample size can be accomplished with only a few modifications to the prior specification. These modifications provide a versatile model with several advantages including the ability to handle both continuous and discrete ordered outcomes and estimation of the full conditional CDF, along with quantiles and other functionals using a single model fit. Inference is based on posterior probability statements and does not require asymptotic assumptions. In addition, the CPM does not require specification of a transformation to meet distributional assumptions since the transformation is estimated nonparametrically. As a result its parameter estimates are invariant to monotonic transformations of the data.

Our implementation of a Bayesian CPM performed reasonably well for the simple simulation scenarios considered. However, the model can produce biased estimates
for quantiles far from the median and conditional quantities further from the model where \(\boldsymbol{X}=0\) and this bias can be exacerbated by censoring.
The model seems best suited for cases when the data are fairly dense and are sufficient to describe the posterior CDF well. In our simulations, a sample size of 50 or 100 was required for reasonably unbiased estimates of parameters and other posterior quantities. The choice of Dirichlet prior concentration with magnitude \(\boldsymbol{\alpha} \approx \frac{1}{J}\) has minimal impact on the bias of posterior estimates, except with small sample sizes. Much larger concentration parameters (e.g., \(\boldsymbol{\alpha}=1/2\)) may be too informative. As with all Bayesian models estimated with MCMC, checks of model convergence, model fit, and the posterior distribution are important. This is especially true when modeling a mixed continuous/discrete or when interest lies in quantities conditional on covariates far from the observed mean values.

Finally, there are several of limitations of the current model that present an opportunity for improvement. First, the number of distinct outcome values is assumed to be known \emph{a priori}, that is we condition on \(J\) categories. In practice, the number of distinct continuous outcome values is unlikely to be available before data collection, so the prior cannot be specified without reference to the observed data. Relatedly, because the number of categories is fixed, the model cannot accommodate new observations for an unobserved category; once the initial prior is set, there is no way to add categories and all predictions are assumed to fall into one of the original categories. It may be possible to overcome this limitation by substituting the Dirichlet prior for a infinite-dimensional Bayesian nonparametric analog, such as a Dirichlet process prior, at the expense of additional complexity and computation time. Next, the choice of link function, and the implied error distribution on the scale of the latent untransformed data is also assumed to be known. If primary interest is not inference for the parameters, specification of the link could be avoided by either estimating the link nonparametrically, although other assumptions may be required for identifiability {[}\protect\hyperlink{ref-song_semiparametric_2012}{18}--\protect\hyperlink{ref-mallick_bayesian_2003}{20}{]}, or using a more flexible mixture link function {[}\protect\hyperlink{ref-lang_bayesian_1999}{8}{]}.

\newpage

\hypertarget{acknowledgments}{%
\subsection{Acknowledgments}\label{acknowledgments}}

We would like to thank Dr.~John Koethe for providing the biomarker data and Yuqi Tian and Dr.~Chun Li for helpful comments and review of early versions of this work. This study was supported by funding from United States National Institutes of Health (R01AI093234, P30AI110527, K23100700, K23AT002508, P30AI54999, and UL1TR000445). This project was also supported in part by an appointment to the Research Participation Program at the Office of Biostatistics, Center for Drug Evaluation and Research, U.S. Food and Drug Administration, administered by the Oak Ridge Institute for Science and Education through an interagency agreement between the U.S. Department of Energy and FDA.

\hypertarget{author-contributions}{%
\subsubsection{Author contributions}\label{author-contributions}}

Study conception and design: NTJ, FEH. Analyses: NTJ. Drafting manuscript: NTJ. Critical reading of manuscript and edits: NTJ, FEH, BES

\newpage

\hypertarget{references}{%
\section*{References}\label{references}}
\addcontentsline{toc}{section}{References}

\hypertarget{refs}{}
\begin{CSLReferences}{0}{0}
\leavevmode\hypertarget{ref-agresti_categorical_2002}{}%
\CSLLeftMargin{1. }
\CSLRightInline{Agresti A. Categorical {Data} {Analysis}. 2nd ed. New York: Wiley-Interscience; 2002. }

\leavevmode\hypertarget{ref-walker_estimation_1967}{}%
\CSLLeftMargin{2. }
\CSLRightInline{Walker SH, Duncan DB. Estimation of the probability of an event as a function of several independent variables. Biometrika {[}Internet{]} 1967 {[}cited 2020 Jul 17{]};54:167--79. Available from: \url{https://academic.oup.com/biomet/article-lookup/doi/10.1093/biomet/54.1-2.167}}

\leavevmode\hypertarget{ref-peter_mccullagh_regression_1980}{}%
\CSLLeftMargin{3. }
\CSLRightInline{McCullagh P. Regression {Models} for {Ordinal} {Data}. Journal of the Royal Statistical Society. Series B (Methodological) {[}Internet{]} 1980;42:109--42. Available from: \url{http://www.jstor.org/stable/2984952}}

\leavevmode\hypertarget{ref-albert_bayesian_1993}{}%
\CSLLeftMargin{4. }
\CSLRightInline{Albert JH, Chib S. Bayesian {Analysis} of {Binary} and {Polychotomous} {Response} {Data}. Journal of the American Statistical Association {[}Internet{]} 1993 {[}cited 2018 May 9{]};88:669. Available from: \url{https://www.jstor.org/stable/2290350?origin=crossref}}

\leavevmode\hypertarget{ref-albert_bayesian_1997}{}%
\CSLLeftMargin{5. }
\CSLRightInline{Albert J, Chib S. Bayesian {Methods} for {Cumulative}, {Sequential} and {Two}-step {Ordinal} {Data} {Regression} {Models}. 1997;33. }

\leavevmode\hypertarget{ref-johnson_ordinal_1999}{}%
\CSLLeftMargin{6. }
\CSLRightInline{Johnson VE, Albert J. Ordinal data modeling. New York: Springer; 1999. }

\leavevmode\hypertarget{ref-peterson_partial_1990}{}%
\CSLLeftMargin{7. }
\CSLRightInline{Peterson B, Harrell FE. Partial {Proportional} {Odds} {Models} for {Ordinal} {Response} {Variables}. Applied Statistics {[}Internet{]} 1990 {[}cited 2020 Jul 18{]};39:205. Available from: \url{https://www.jstor.org/stable/10.2307/2347760?origin=crossref}}

\leavevmode\hypertarget{ref-lang_bayesian_1999}{}%
\CSLLeftMargin{8. }
\CSLRightInline{Lang JB. Bayesian ordinal and binary regression models with a parametric family of mixture links. Computational Statistics \& Data Analysis {[}Internet{]} 1999 {[}cited 2018 Jul 17{]};31:59--87. Available from: \url{http://linkinghub.elsevier.com/retrieve/pii/S0167947399000079}}

\leavevmode\hypertarget{ref-congdon_bayesian_2005}{}%
\CSLLeftMargin{9. }
\CSLRightInline{Congdon P. Bayesian models for categorical data. Chichester ; New York: Wiley; 2005. }

\leavevmode\hypertarget{ref-liu_modeling_2017}{}%
\CSLLeftMargin{10. }
\CSLRightInline{Liu Q, Shepherd BE, Li C, Harrell FE. Modeling continuous response variables using ordinal regression. Statistics in Medicine {[}Internet{]} 2017 {[}cited 2018 Jan 19{]};36:4316--35. Available from: \url{http://onlinelibrary.wiley.com/doi/10.1002/sim.7433/abstract}}

\leavevmode\hypertarget{ref-harrell_regression_2015}{}%
\CSLLeftMargin{11. }
\CSLRightInline{Harrell FE. Regression modeling strategies: With applications to linear models, logistic and ordinal regression, and survival analysis. Second edition. Cham Heidelberg New York: Springer; 2015. }

\leavevmode\hypertarget{ref-tian_empirical_2019}{}%
\CSLLeftMargin{12. }
\CSLRightInline{Tian Y, Hothorn T, Li C, Harrell FE, Shepherd BE. An empirical comparison of two novel transformation models. Statistics in Medicine {[}Internet{]} 2019 {[}cited 2020 Feb 7{]};Available from: \url{https://onlinelibrary.wiley.com/doi/abs/10.1002/sim.8425}}

\leavevmode\hypertarget{ref-zeng_maximum_2007}{}%
\CSLLeftMargin{13. }
\CSLRightInline{Zeng D, Lin DY. Maximum likelihood estimation in semiparametric regression models with censored data. Journal of the Royal Statistical Society: Series B (Statistical Methodology) {[}Internet{]} 2007 {[}cited 2020 Aug 4{]};69:507--64. Available from: \url{https://rss.onlinelibrary.wiley.com/doi/abs/10.1111/j.1369-7412.2007.00606.x}}

\leavevmode\hypertarget{ref-gelfand_approaches_1999}{}%
\CSLLeftMargin{14. }
\CSLRightInline{Gelfand AE. Approaches for {Semiparametric} {Bayesian} {Regression}. In: Ghosh S, editor. Asymptotics, {Nonparametrics}, and {Time} {Series}. CRC Press; 1999. page 615--38.}

\leavevmode\hypertarget{ref-brunner_bayesian_1995}{}%
\CSLLeftMargin{15. }
\CSLRightInline{Brunner LJ. Bayesian linear regression with error terms that have symmetric unimodal densities. Journal of Nonparametric Statistics {[}Internet{]} 1995 {[}cited 2020 Aug 13{]};4:335--48. Available from: \url{http://www.tandfonline.com/doi/abs/10.1080/10485259508832625}}

\leavevmode\hypertarget{ref-kottas_bayesian_2001}{}%
\CSLLeftMargin{16. }
\CSLRightInline{Kottas A, Gelfand AE. Bayesian {Semiparametric} {Median} {Regression} {Modeling}. Journal of the American Statistical Association {[}Internet{]} 2001 {[}cited 2020 Jul 28{]};96:1458--68. Available from: \url{http://www.tandfonline.com/doi/abs/10.1198/016214501753382363}}

\leavevmode\hypertarget{ref-deyoreo_bayesian_2020}{}%
\CSLLeftMargin{17. }
\CSLRightInline{DeYoreo M, Kottas A. Bayesian nonparametric density regression for ordinal responses. In: Flexible {Bayesian} regression modelling. Academic Press; 2020. page 65--89.}

\leavevmode\hypertarget{ref-song_semiparametric_2012}{}%
\CSLLeftMargin{18. }
\CSLRightInline{Song X-Y, Lu Z-H. Semiparametric transformation models with {Bayesian} {P}-splines. Statistics and Computing {[}Internet{]} 2012 {[}cited 2020 Jul 2{]};22:1085--98. Available from: \url{http://link.springer.com/10.1007/s11222-011-9280-x}}

\leavevmode\hypertarget{ref-tang_semiparametric_2018}{}%
\CSLLeftMargin{19. }
\CSLRightInline{Tang N, Wu Y, Chen D. Semiparametric {Bayesian} analysis of transformation linear mixed models. Journal of Multivariate Analysis {[}Internet{]} 2018 {[}cited 2020 Aug 1{]};166:225--40. Available from: \url{https://linkinghub.elsevier.com/retrieve/pii/S0047259X18300976}}

\leavevmode\hypertarget{ref-mallick_bayesian_2003}{}%
\CSLLeftMargin{20. }
\CSLRightInline{Mallick BK, Walker S. A {Bayesian} semiparametric transformation model incorporating frailties. Journal of Statistical Planning and Inference {[}Internet{]} 2003 {[}cited 2020 Jun 30{]};112:159--74. Available from: \url{https://linkinghub.elsevier.com/retrieve/pii/S0378375802003300}}

\leavevmode\hypertarget{ref-lin_semiparametric_2012}{}%
\CSLLeftMargin{21. }
\CSLRightInline{Lin J, Sinha D, Lipsitz S, Polpo A. Semiparametric {Bayesian} {Survival} {Analysis} using {Models} with {Log}-{Linear} {Median}. Biometrics {[}Internet{]} 2012 {[}cited 2020 Aug 4{]};68:1136--45. Available from: \url{https://www.jstor.org/stable/41806032}}

\leavevmode\hypertarget{ref-damien_surviving_2013}{}%
\CSLLeftMargin{22. }
\CSLRightInline{Hanson TE, Jara A. Surviving fully {Bayesian} nonparametric regression models {[}Internet{]}. In: Damien P, Dellaportas P, Polson NG, Stephens DA, editors. Bayesian {Theory} and {Applications}. Oxford University Press; 2013 {[}cited 2020 Jul 2{]}. page 593--616.Available from: \url{http://www.oxfordscholarship.com/view/10.1093/acprof:oso/9780199695607.001.0001/acprof-9780199695607-chapter-30}}

\leavevmode\hypertarget{ref-hanson_bayesian_2007}{}%
\CSLLeftMargin{23. }
\CSLRightInline{Hanson T, Yang M. Bayesian {Semiparametric} {Proportional} {Odds} {Models}. Biometrics {[}Internet{]} 2007 {[}cited 2020 Jun 30{]};63:88--95. Available from: \url{http://onlinelibrary.wiley.com/doi/abs/10.1111/j.1541-0420.2006.00671.x}}

\leavevmode\hypertarget{ref-ibrahim_bayesian_2010}{}%
\CSLLeftMargin{24. }
\CSLRightInline{Ibrahim JG, Chen M-H, Sinha D. Bayesian survival analysis. Softcover repr. of the hardcover 1st edition 2001, corr. 2nd printing. New York: Springer; 2010. }

\leavevmode\hypertarget{ref-muller_bayesian_2015}{}%
\CSLLeftMargin{25. }
\CSLRightInline{Müller P, Quintana FA, Jara A, Hanson T. Bayesian nonparametric data analysis. Cham: Springer; 2015. }

\leavevmode\hypertarget{ref-hjort_bayesian_2010}{}%
\CSLLeftMargin{26. }
\CSLRightInline{Hjort NL, editor. Bayesian nonparametrics. Cambridge, UK ; New York: Cambridge University Press; 2010. }

\leavevmode\hypertarget{ref-mckinley_bayesian_2015}{}%
\CSLLeftMargin{27. }
\CSLRightInline{McKinley TJ, Morters M, Wood JLN. Bayesian {Model} {Choice} in {Cumulative} {Link} {Ordinal} {Regression} {Models}. Bayesian Analysis {[}Internet{]} 2015 {[}cited 2019 Nov 2{]};10:1--30. Available from: \url{http://arxiv.org/abs/1503.07642}}

\leavevmode\hypertarget{ref-betancourt_ordinal_2019}{}%
\CSLLeftMargin{28. }
\CSLRightInline{Betancourt M. Ordinal {Regression} {[}Internet{]}. 2019 {[}cited 2020 Jul 3{]};Available from: \url{https://betanalpha.github.io/assets/case_studies/ordinal_regression.html}}

\leavevmode\hypertarget{ref-gelman_bayesian_2014}{}%
\CSLLeftMargin{29. }
\CSLRightInline{Gelman A, Carlin J, Stern H, Dunson D, Vehtari A, Rubin D. Bayesian {Data} {Analysis}. Third edition. Boca Raton: CRC Press; 2014. }

\leavevmode\hypertarget{ref-berger_overall_2015}{}%
\CSLLeftMargin{30. }
\CSLRightInline{Berger JO, Bernardo JM, Sun D. Overall {Objective} {Priors}. Bayesian Analysis {[}Internet{]} 2015 {[}cited 2020 Jul 2{]};10:189--221. Available from: \url{http://projecteuclid.org/euclid.ba/1422556416}}

\leavevmode\hypertarget{ref-stan_development_team_rstan:_2018}{}%
\CSLLeftMargin{31. }
\CSLRightInline{Team SD. {RStan}: The {R} interface to {Stan} {[}Internet{]}. 2018;Available from: \url{http://mc-stan.org/}}

\leavevmode\hypertarget{ref-neal_mcmc_2011}{}%
\CSLLeftMargin{32. }
\CSLRightInline{Neal R. {MCMC} using {Hamiltonian} {Dynamics}. In: Fitzmaurice G, Brooks S, Gelman A, Jones GL, Meng X-L, editors. Handbook of {Markov} {Chain} {Monte} {Carlo}. New York: CRC Press, Taylor \& Francis Group; 2011. page 113--62.}

\leavevmode\hypertarget{ref-koethe_serum_2012}{}%
\CSLLeftMargin{33. }
\CSLRightInline{Koethe JR, Bian A, Shintani AK, Boger MS, Mitchell VJ, Erdem H, et al. Serum {Leptin} {Level} {Mediates} the {Association} of {Body} {Composition} and {Serum} {C}-{Reactive} {Protein} in {HIV}-{Infected} {Persons} on {Antiretroviral} {Therapy}. AIDS Research and Human Retroviruses {[}Internet{]} 2012 {[}cited 2020 Aug 13{]};28:552--7. Available from: \url{http://www.liebertpub.com/doi/10.1089/aid.2011.0232}}

\leavevmode\hypertarget{ref-koethe_metabolic_2015}{}%
\CSLLeftMargin{34. }
\CSLRightInline{Koethe JR, Grome H, Jenkins CA, Kalams SA, Sterling TR. The metabolic and cardiovascular consequences of obesity in persons with {HIV} on long-term antiretroviral therapy: AIDS {[}Internet{]} 2015 {[}cited 2020 Aug 13{]};1. Available from: \url{http://journals.lww.com/00002030-900000000-97959}}

\leavevmode\hypertarget{ref-vehtari_practical_2017}{}%
\CSLLeftMargin{35. }
\CSLRightInline{Vehtari A, Gelman A, Gabry J. Practical {Bayesian} model evaluation using leave-one-out cross-validation and {WAIC}. Statistics and Computing {[}Internet{]} 2017 {[}cited 2019 Jun 25{]};27:1413--32. Available from: \url{http://link.springer.com/10.1007/s11222-016-9696-4}}

\leavevmode\hypertarget{ref-stern_bayesian_2005}{}%
\CSLLeftMargin{36. }
\CSLRightInline{Stern HS, Sinharay S. Bayesian {Model} {Checking} and {Model} {Diagnostics} {[}Internet{]}. In: Handbook of {Statistics}. Elsevier; 2005 {[}cited 2020 Aug 10{]}. page 171--92.Available from: \url{https://linkinghub.elsevier.com/retrieve/pii/S016971610525006X}}

\end{CSLReferences}

\newpage

\hypertarget{supplemental-material}{%
\section*{Supplemental Material}\label{supplemental-material}}
\addcontentsline{toc}{section}{Supplemental Material}

\beginsupplement

\end{document}